\documentclass[aps,prc,twocolumn,showpacs,preprintnumbers,
               nofootinbib,float,superscriptaddress,longbibliography]{revtex4-1}

\usepackage{graphics}      
\usepackage{graphicx}      
\usepackage{longtable}     
\usepackage{url}           
\usepackage{bm}            
\usepackage{amsmath}
\usepackage{amssymb}
\usepackage{float}
\usepackage{physics}
\usepackage{amsmath,amssymb}
\usepackage[colorlinks=true, pdfstartview=FitV, linkcolor=red, citecolor=blue, urlcolor=blue]{hyperref}
\usepackage{color}
\usepackage[utf8]{inputenc}

\usepackage{capt-of}

\begin{document}

\title{Diffractive dijet production in impact parameter dependent saturation models}

\author{Farid Salazar}
\email{farid.salazarwong@stonybrook.edu}
\affiliation{\small Physics Department, Brookhaven National Laboratory, Bldg. 510A, Upton, NY 11973, USA}
\affiliation{\small Department of Physics and Astronomy, Stony Brook University, Stony Brook, New York 11794, USA }

\author{Bj\"orn Schenke}
\email{bschenke@bnl.gov}
\affiliation{\small Physics Department, Brookhaven National Laboratory, Bldg. 510A, Upton, NY 11973, USA}

\date{\today}

\begin{abstract}

We study coherent diffractive dijet production in electron-hadron and electron-nucleus collisions within the dipole picture. We provide semi-analytic results for the differential cross section and elliptic anisotropy in the angle between mean dijet transverse momentum and hadron recoil momentum. We demonstrate the direct relation between angular moments of the dipole amplitude in coordinate space and angular moments of the diffractive dijet cross sections.
To perform explicit calculations we employ two different saturation models, extended to include the target geometry.
In the limit of large photon virtuality or quark masses, we find fully analytic results that allow direct insight into how the differential cross section and elliptic anisotropy depend on the saturation scale, target geometry, and kinematic variables.
We further provide numerical results for more general kinematics in collisions at a future electron-ion collider, and study the effects of approaching the saturated regime on diffractive dijet observables.

\end{abstract}

\maketitle

\section{Introduction}

Exclusive dijet production in coherent diffractive processes in electron-nucleus (and electron-hadron) scattering provides important information on the structure of the target in both coordinate and momentum space. In certain kinematic limits it was shown \cite{Hatta:2016dxp,Hagiwara:2017fye} that diffractive dijet production cross sections can be directly related to the 5-dimensional gluon Wigner distribution of the target \cite{Ji:2003ak,Belitsky:2003nz}.
This can be used to constrain both generalized parton distributions (GPDs) \cite{Diehl:2003ny,Belitsky:2005qn} and transverse momentum dependent parton distributions (TMDs) \cite{Collins:1981uw,Mulders:2000sh,Meissner:2007rx}.

Recently, diffractive dijet production in photon-hadron collisions was studied in the dipole picture \cite{Altinoluk:2015dpi,Mantysaari:2019csc}, which is an appropriate framework for high energy collisions.
Interesting structures, such as diffractive dips, were found in the differential cross sections as functions of the hadron recoil momentum and the mean dijet transverse momentum. Furthermore, the dependence of the cross section on the angle between those two momenta exhibited quite a complex behavior with the sign and magnitude of the elliptic anisotropy coefficient depending on the photon polarization, virtuality of the photon, mean dijet transverse momentum, and details of the dipole model.

In this work we aim to provide analytic insight into what physical properties of the projectile and target are responsible for the observed features in the angle averaged and angular dependent cross sections, and how these features depend on the kinematic variables of the studied process. For this purpose we concentrate on relatively simple dipole models, which we extend to include a spatial dependence of the target's color charge density, and, if necessary in the model, an additional explicit correlation between the dipole orientation and the impact parameter of the collision. These features are necessary to obtain non-trivial results for the diffractive cross sections and their angular dependence.

In particular, we first study the Golec-Biernat Wusthoff (GBW) model \cite{GolecBiernat:1998js,GolecBiernat:1999qd}, extended to include a spatially dependent target and angular correlations. In the limit of large photon virtuality $Q^2$ and/or mass of the (anti-)quark in the dijet, we evaluate the diffractive dijet production cross sections analytically, which allows for important insight into how projectile and target properties affect the experimental observables. 

A somewhat more realistic description, in particular of the angular dependence, can be achieved by using an impact parameter dependent McLerran Venugopalan (IP-MV) model \cite{McLerran:1993ni,McLerran:1993ka,McLerran:1994vd}, where the angular modulation of the dipole amplitude depends on spatial gradients of the transverse color charge distribution, as previously shown in \cite{Iancu:2017fzn}. We discuss how the results are modified from the GBW model in the limit of large $Q^2$ and/or mass of the (anti-)quark in the dijet. However, no exact analytical result can be obtained within this model, even with the simplification of the studied limit. 
We thus evaluate the diffractive dijet cross sections in this model numerically, which also reveals interesting features and their dependence on the target and projectile wave function, as well as the kinematic variables. We note that more complex saturation models, such as IPSat \cite{Kowalski:2003hm,Kowalski:2008sa} and b-CGC \cite{Kowalski:2006hc,Watt:2007nr} also include impact parameter dependence, but for the sake of simplicity and being able to find analytic expressions, we stick to the two simpler models mentioned above.

Finally, by varying the saturation scale and species of the target nucleus, we analyze the effects of approaching the saturated regime on the diffractive dijet observables. We find a clear nuclear dependence on the growth of the cross section with decreasing $x$, as well as a characteristic dip structure of the differential cross section as a function of recoil momentum, which varies with changing saturation scale. The latter is a consequence of approaching the black disk limit, which means that the effective spatial color charge distribution of the proton deviates increasingly from a Gaussian with decreasing $x$, because of unitarity constraints. 

This work should provide important guidance for future experiments at an electron ion collider (EIC) \cite{Accardi:2012qut,Aschenauer:2017jsk}, where the spatial and momentum structure of nuclear targets can be probed to unprecedented precision. 

We note that geometrical effects, similar to the ones appearing in this work, have also been studied for the case of elliptic anisotropies of gluon production in A+A \cite{Teaney:2002kn} and p+A (p+p) \cite{Iancu:2017fzn,Kovner:2018fxj} collisions. However, such processes, as well as other, like inclusive dijet production in e+p collisions \cite{Dominguez:2011wm,Dumitru:2018kuw} will have other sources of anisotropy, dominantly from quantum interference effects.
In contrast, for the case of coherent diffractive processes as studied in this work, the geometric effects are the dominant source of anisotropy. 

This paper is organized as follows: In Section \ref{sec:cohdiffdipoles} we set up the notation and review the basic ingredients of coherent diffractive dijet production in the dipole framework. 

In Section \ref{sec:semi-analytic} we derive a semi-analytic expression for coherent diffractive dijet differential cross sections from the dipole amplitude. In Section \ref{sec:semianalytic_noangular} we obtain expressions for the case of a dipole amplitude without angular correlations. We then extend the analysis to dipole amplitudes with angular dependence, and derive formulas for the differential cross section and elliptic anisotropy in Section \ref{sec:semianalytic_angular} .

In Section \ref{sec:dipoleModels} we review the models used to specify the dipole amplitude, and in Section \ref{sec:analytics} we investigate analytic properties of the differential cross section and elliptic anisotropy for the two models.

In Section \ref{sec:numrel} we present numerical results for the coherent diffractive dijet cross sections and their angular modulation in the IP-MV model. We present both the dependence on the dijet momentum $P$ (Section \ref{sec:Pdep}) and on the recoil momentum $\Delta$ (Section \ref{sec:Deltadep}). In both cases, we study proton and gold targets. We consider different photon virtualities and saturation scales expected to be reached at a future EIC. In Section \ref{sec:saturatedregime} we discuss observable effects of approaching the saturated regime, focusing on the differential cross section. 

We summarize our results and discuss the limitations of our analysis and possible extensions of these techniques to the study of other processes in Section \ref{sec:conclusions}. We include multiple appendices with technical details of the calculations.

\section{Coherent diffractive dijet production in the dipole picture}
\label{sec:cohdiffdipoles}

\subsection{The dipole picture}
In the dipole picture, coherent diffractive dijet production in electron-nucleus (-proton) scattering can be described as a two step process: the fluctuation of the virtual photon (emitted by the electron) into a color neutral quark -- anti-quark dipole and the scattering of the dipole with the target (proton or nucleus). In the Color Glass Condensate (CGC) framework at high energies (small Bjorken-$x$) the target is described by strong classical color fields generated by partons at larger $x$, which are static and localized color sources. The saturation of these fields in transverse momentum space is characterized by the saturation scale $Q_s$ which grows with decreasing $x$.
We note that our calculations are performed at leading order. Next-to-leading order (NLO) impact factors for both diffractive dijet and vector meson production have been derived \cite{Boussarie:2016ogo,Boussarie:2016bkq}. Together with next-to leading order corrections to the CGC evolution kernel \cite{Balitsky:2008zza,Balitsky:2013fea,Grabovsky:2013mba,Kovner:2013ona,Balitsky:2014mca,Caron-Huot:2015bja,Lublinsky:2016meo}, complete NLO
calculations will be possible. 

The fluctuation of the photon with virtuality $Q^2$ into a quark -- anti-quark dipole of size $r$ can be computed in quantum electrodynamics and is described by the light-cone wave functions \cite{Bjorken:1970ah,Nikolaev:1990ja,Dosch:1996ss,Beuf:2011xd,Kovchegov:2012mbw}:

\begin{align}
\Psi_L(\bm{r}) =  \frac{e Z_f}{2\pi} (z_0 z_1)^{3/2} \delta_{\beta_0 \beta_1}  2 Q (1-\delta_{\sigma_0 \sigma_1}) K_0\left( \varepsilon_f r  \right)\,, \label{PsiL} 
\end{align}
\begin{align}
\Psi_{T \lambda}(\bm{r}) =&\frac{e Z_f}{2\pi} \sqrt{z_0 z_1} \delta_{\beta_0 \beta_1}    \notag\\ &\times\bigg[ \delta_{\sigma_0 \sigma_1} \frac{m_f}{\sqrt{2}} (1+ \sigma_0 \lambda) K_0(\varepsilon_f r)  \nonumber \\ & ~~~~ +  (1-\delta_{\sigma_0 \sigma_1}) (z_1 -z_0 - \sigma_0 \lambda) \notag\\ & ~~~~ ~~~~\times i \varepsilon_f \frac{\bm{\epsilon}_\lambda \cdot \bm{r}}{r}  K_1 ( \varepsilon_f r ) \bigg] \label{PsiT}\,,
\end{align}
where $\varepsilon^2_f = z_0 z_1 Q^2 + m_f^2$
and $z_0$ and $z_1=1-z_0$ are the fractions of the longitudinal photon momentum carried by the quark and anti-quark, respectively. The subscript $\sigma_i \in \{-1,1\}$ refers to the helicity and $\beta_j$ to the color of the quark (anti-quark) ($i,j \in \{0,1\}$), and $m_f$ and $e Z_f$ are the mass and electric charge of the quark of flavor $f$. The two dimensional vector $\bm{\epsilon}_\lambda$ denotes the transverse polarization, with $\lambda=\pm 1$.

The functions $K_0(z)$ and $K_1(z)$ are modified Bessel functions of the second kind, which decay rapidly as their arguments increase. For massive quarks or large photon virtualities, this implies that only small dipoles will contribute to the process to be studied, which is favorable since large dipole contributions are affected by uncontrolled infrared physics at the confinement scale \cite{Mantysaari:2019csc}. Consequently, we will restrict our analysis to charm quarks.

The scattering of the dipole off the strong classical color fields is treated within the eikonal approximation, in which the transverse coordinates of the dipole are unchanged as it travels through the target, and the scattering is characterized by the color rotation of the quark and anti-quark. The color rotation of the quark via multiple scattering is encoded in the longitudinal Wilson line in the fundamental representation of $SU(N_c=3)$
\begin{align}
V^\dagger(\bm{x}) = e^{ig \int dz^- A^{+,a}(z^-,\bm{x})t^a}\,. \label{WilsonLine}
\end{align} 
Therefore, the scattering amplitude takes a simple form in transverse coordinate space (because the dipole is initially in a color singlet state, we use the color delta function $\delta_{\beta_0 \beta_1}$ to contract the inner indices of the longitudinal Wilson lines to form the product)
\begin{align}
\mathcal{S}(\bm{x_0},\bm{x_1})_{ij} = \left[ V^\dagger(\bm{x_0}) V(\bm{x_1}) \right]_{ij},
\end{align}
where $\bm{x}_0$ and $\bm{x_1}$  are the transverse coordinates, and $i$ and $j$ are the color indices of the quark and anti-quark after scattering off the target, respectively. 

The interaction of the virtual photon (with given polarization) with the proton or nucleus is given by the scattering matrix element 
\begin{align}
\mathcal{M}_{ij}(\bm{x}_0,\bm{x}_1)_{T \lambda, L} =& \Psi_{T \lambda, L} ( \bm{x}_0-\bm{x}_1)\notag\\ &\times\left(\mathcal{I}_{ij} - \mathcal{S}_{ij} (\bm{x}_0,\bm{x}_1) \right)\,, \label{AmplitudeCoordinateSpace}
\end{align}
where $\mathcal{I}_{ij}$ is the unit matrix.

The non-perturbative information about the degrees of freedom of the target is encoded in the longitudinal Wilson lines.

In the next section we focus on the specific case of coherent diffractive processes, which require certain constraints on the color structure of the matrix element and the procedure for averaging over the target's color charge densities.

\subsection{From dipole amplitude to coherent diffractive dijet cross section}\label{sec:dipoletoxsec}

Before we discuss how to obtain the differential cross section from the dipole amplitude, it is important to clarify the terminology regarding our process of interest. Diffractive refers to the absence of net color exchange between the dipole and the target during the scattering; i.e., the color rotation of the quark compensates that of the anti-quark. A rapidity gap is the experimental signature for this color singlet final state, as no color string is formed between dipole and target, whose breaking would lead to particle production at intermediate rapidities. We ensure a color singlet final state by restricting to the diagonal in color space.

Coherent refers to scattering processes in which the target remains intact. Coherent processes require that the average over color charge densities is taken at the level of the amplitude. This differs from inclusive processes in which the average is taken at the level of the cross section \cite{Miettinen:1978jb,Kovner:2001vi,Kovchegov:1999kx}. 

Therefore, the object of interest in coherent diffractive dijet production is given by the color-diagonal averaged matrix element:
\begin{align}
 \left \langle \mathcal{M}_{ij}(\bm{x}_0,\bm{x}_1)_{T \lambda, L} \right \rangle =\Psi_{T \lambda, L} ( \bm{x}_0-\bm{x}_1)
    D(\bm{x}_0,\bm{x}_1) \delta_{ij}
\end{align}
For the sake of simplicity, we will denote the piece in front of $\delta_{ij}$ as $\mathcal{M}(\bm{x}_0,\bm{x}_1)_{T \lambda, L}$. The dipole amplitude is defined as
\begin{align}
D(\bm{x}_0,\bm{x}_1) = 1 - \frac{1}{3}\left \langle \tr\left( V^\dagger(\bm{x_0}) V(\bm{x_1} \right) \right \rangle \,,
\end{align}
and contains information on the spatial structure of the target.

In order to compute the cross section we express the amplitude in momentum space
\begin{align}
\langle \tilde{\mathcal{M}}_{ij}&(\bm{p}_0,\bm{p}_1)_{T\lambda,L} \rangle  = \notag\\ & \hspace{-0.2cm}\int d^2 \bm{x}_0 d^2 \bm{x}_1 e^{-i \bm{p}_0 \cdot \bm{x}_0} e^{-i \bm{p}_1 \cdot \bm{x}_1} \mathcal{M}_{T\lambda,L}(\bm{x}_0,\bm{x}_1) \delta_{ij} \,,
\end{align}
where $\bm{p}_0$ and $\bm{p}_1$ are the transverse momenta of quark jet and anti-quark jet, respectively. 

Before presenting the differential cross section for coherent diffractive dijet production, it is useful to introduce the set of coordinates
\begin{align}\label{rbcoordinates}
\bm{r} &= \bm{x}_0 - \bm{x}_1\,, \nonumber \\
\bm{b} &=\frac{1}{2}\left( \bm{x}_0 + \bm{x}_1 \right)\,,
\end{align}
the dipole and impact parameter vectors, respectively. Their conjugates are
\begin{align}
\bm{P} &= \frac{1}{2}\left( \bm{p}_0 - \bm{p}_1 \right) \,, \nonumber \\
\bm{\Delta} &= \bm{p}_0 + \bm{p}_1 \,,
\end{align}
characterizing the mean dijet transverse momentum and momentum transfer (or nucleus recoil momentum), respectively.

In the frame in which the virtual photon has large longitudinal momentum $q^-$, the differential cross section for coherent diffractive dijet production in electron-nucleus (-proton) scattering is then given by 
\begin{align}
\frac{d \sigma_{T\lambda,L}}{d \Omega} = 
 \frac{N_c \abs{q^-} \delta(q^--p^-_0 -p^-_1)}{ 2(2\pi)^5} \left|  \langle \tilde{\mathcal{M}}_{T\lambda,L} \rangle \right|^2 \label{CrossSection}\,,
\end{align}
where $d \Omega =(d p_0^-/p_0^-)(d p_1^-/p_1^-)  d^2 \bm{P} d^2 \bm{\Delta} $, the longitudinal momenta for quark and anti-quark jet are given by $p^-_0$  and $p^-_1$, respectively.

The averaged amplitude is given by \begin{align}
\langle \tilde{\mathcal{M}}_{T\lambda,L} \rangle = \int d^2 \bm{r}\  d^2 \bm{b} \ e^{-i \bm{P} \cdot \bm{r}}  e^{-i \bm{\Delta} \cdot \bm{b}} \ \Psi_{T \lambda, L} (\bm{r}) D(\bm{r},\bm{b})\,.
\end{align} 
Here we defined
\begin{equation}
    D(\mathbf{r}, \mathbf{b}) = 1 - \frac{1}{3}\left \langle \tr\left( V^\dagger(\mathbf{b}+\mathbf{r}/2) V(\mathbf{b}-\mathbf{r}/2 \right) \right \rangle \,.
\end{equation}

Combining above results, we arrive at expressions for the differential cross section for coherent diffractive dijet production for longitudinally and transversely polarized photons. We sum over helicities and colors (and for the transverse case over the two possible polarizations $\lambda$):

\begin{align}
\frac{d \sigma_{L}}{d \Omega} = &\frac{8 N_c \alpha_{EM}}{(2\pi)^6}  Z^2_f Q^2 z_0^3 z_1^3 \notag \\ & ~~ \times \delta(1-z_0-z_1)  \left | \tilde{F}(\bm{P}, \bm{\Delta}) \right |^2 \label{CrossPhotonL} \,, \\
 \frac{d \sigma_{T}}{d \Omega}  = &\frac{2 N_c \alpha_{EM}}{(2\pi)^6}  Z^2_f z_0 z_1 \delta(1-z_0-z_1) \nonumber \\  & \times \bigg[  \varepsilon_f^2  \zeta^2  \left | \partial_{\bm{P}} \tilde{G}(\bm{P}, \bm{\Delta}) \right |^2  \notag \\ & ~~~~~~ + m_f^2 \left | \tilde{F}(\bm{P}, \bm{\Delta})\right |^2 \bigg] \label{CrossPhotonT} \,.
\end{align}
where we defined $\zeta^2 = z_0^2 + z_1^2$.

The functions $\tilde{F}(\bm{P},\bm{\Delta})$ and $\tilde{G}(\bm{P},\bm{\Delta})$ are given by
\begin{align}
\tilde{F}(\bm{P}, \bm{\Delta}) &= \hspace{-0.12cm}\int d^2\bm{r} d^2 \bm{b} e^{-i \bm{P} \cdot \bm{r}}  e^{-i \bm{\Delta} \cdot \bm{b}} K_0(\epsilon_f r)  D(\bm{r},\bm{b}) \label{F}, \\
\tilde{G}(\bm{P},\bm{\Delta}) & = \hspace{-0.12cm} \int  d^2 \bm{r} d^2 \bm{b} e^{-i \bm{P} \cdot \bm{r}}  e^{-i \bm{\Delta} \cdot \bm{b}} \frac{K_1(\epsilon_f r)}{r}   D(\bm{r},\bm{b}) \label{G}.
\end{align}

These functions depend on the details of the light-cone wave functions and  the dipole amplitude. 

\section{Relation between modes of the dipole amplitude and the diffractive dijet cross section}
\label{sec:semi-analytic}

In this section we derive formulas connecting the modes of the angular correlation of the dipole amplitude, and the corresponding modes of the scattering amplitude in momentum space, and the differential dijet cross section. 

If the target is isotropic, the differential cross section will only depend on the magnitudes $P$, $\Delta$, and the relative angle $\theta_{P \Delta} = \theta_P - \theta_\Delta$. The main goal of this work is to describe the dependence of the diffractive dijet differential cross section on these three variables and provide analytic insight into the relation between the target and projectile properties and experimental observables.

In order to study the angular dependence, it is useful to decompose the differential cross section in Fourier modes
\begin{align}\label{eq:Fourierxsec}
\frac{d \sigma_{L(T)}}{ d\Omega} =&\delta(1-z_0-z_1) \times \nonumber \\ &\left( \frac{d \sigma_{L(T),0}}{ d\Omega}   + 2\frac{d \sigma_{L(T),2}}{ d\Omega}  \cos 2\theta_{P \Delta} + ... \right) \,.
\end{align}
The first term $d \sigma_{L(T),0}/ d\Omega$ is the  differential cross section averaged over angle $\theta_{P \Delta}$ (from now on we will refer to it simply as differential cross section). The second term $d \sigma_{L(T),2}/ d \Omega$ is the elliptic anisotropy emerging from angular correlations between $\bm{P}$ and $\bm{\Delta}$.

Any momentum correlations are encoded in the functions $\tilde{F}(\bm{P},\bm{\Delta})$ and $\tilde{G}(\bm{P},\bm{\Delta})$, and they arise from the angular correlations between $\bm{r}$ and $\bm{b}$ in the dipole amplitude via Fourier transform. In order to make the connection between Fourier modes of the dipole amplitude and of the functions $\tilde{F}$ and $\tilde{G}$ more explicit, we decompose the dipole amplitude into Fourier modes:
\begin{align}\label{eq:dipoleExpansion}
    D(\bm{r},\bm{b}) = D_0(r,b) + 2 D_2(r,b) \cos 2 \theta_{rb} + ...
\end{align}
We can then evaluate the modes of  $\tilde{F}(\bm{P},\bm{\Delta})$ and $\tilde{G}(\bm{P},\bm{\Delta})$ using the following mathematical relation:
\begin{align}
\frac{\tilde{F}_k(P,\Delta)}{(2\pi)^2} &= (-1)^k \hspace{-0.1cm}\int rdr  b db J_k(Pr) J_k(\Delta b) F_k(r,b) \,,\label{ModeRelation}
\end{align}
with equivalent expressions for $\tilde{G}$.

This general result shows a one-to-one correspondence between modes of a function and modes of its Fourier transform. The derivation of Eq.\,\eqref{ModeRelation} can be found in Appendix \ref{app:moderelation}.

For our particular definition of $\tilde{F}$ (Eq.\,\eqref{F}) one has
\begin{align}
F_0(r,b) =  K_0(\varepsilon_f r) D_0(r,b) \nonumber \\
F_2(r,b) =  K_0(\varepsilon_f r) D_2(r,b) 
\end{align}
and similar expressions for $G_0$ and $G_2$.

\subsection{Dipole without angular correlations}
\label{sec:semianalytic_noangular}

We first review results in the absence of angular correlations. In this case one has $\tilde{F}(\bm{P},\bm{\Delta}) = \tilde{F}_0(P,\Delta)$ and $\tilde{G}(\bm{P},\bm{\Delta}) = \tilde{G}_0(P,\Delta)$. Therefore, using the relation in Eq.\,\eqref{ModeRelation} for $k=0$, we have
\begin{align}
\frac{\tilde{F}(\bm{P}, \bm{\Delta})}{(2\pi)^2}= \int r dr b d b  J_0(Pr)J_0(\Delta b) K_0(\epsilon_f r) D_0(r,b) \,,\nonumber \\
\frac{\tilde{G}(\bm{P},\bm{\Delta})}{(2 \pi)^2 } = \int rdr bdb  J_0(Pr) J_0(\Delta b) \frac{K_1(\epsilon_f r)}{r} D_0(r,b)\,.
\end{align}

The formulas above combined with Eq.\,\eqref{CrossPhotonL} and Eq.\,\eqref{CrossPhotonT} result in
\begin{align}
\frac{d \sigma_{L}}{d \Omega}  &=   \delta(1-z_0-z_1) \frac{8 N_c \alpha_{EM}}{(2\pi)^2} Z^2_f Q^2 z_0^3 z_1^3\nonumber \\  & \times\left | \int r dr b d b  J_0(Pr)J_0(\Delta b) K_0(\epsilon_f r)  D_0(r,b) \right |^2 \label{CrossL0}\,, 
\end{align}
\begin{align}
&\frac{d \sigma_{T}}{d \Omega}  = \delta(1-z_0-z_1)  \frac{2 N_c \alpha_{EM}}{(2\pi)^2} Z^2_f  z_0 z_1 \nonumber \\ &\times \left\lbrace \varepsilon_f^2 \zeta^2  \left | \int rdr bdb  J_1(Pr) J_0(\Delta b) K_1(\epsilon_f r) D_0(r,b) \right |^2  \right.  \nonumber \\ &\left.   ~~~~ + m_f^2  \left | \int r dr b d b  J_0(Pr)J_0(\Delta b) K_0(\epsilon_f r)  D_0(r,b) \right |^2  \right\rbrace \,, \label{CrossT0}
\end{align}
where in the second expression we have used the identity for the derivative of the Bessel function of the first kind $J'_0(z)=-J_1(z)$.
These expressions have been previously obtained in \cite{Altinoluk:2015dpi}.

\subsection{Dipole with angular correlations}
\label{sec:semianalytic_angular}

We proceed to compute the differential cross sections in the presence of angular correlations in the dipole amplitude. This is one of the main results of our paper. We will find corrections to Eq.\,\eqref{CrossL0} and Eq.\,\eqref{CrossT0}, and most importantly we will derive an expression for the elliptic anisotropy (c.f. Eq.\,\eqref{eq:Fourierxsec}).
We divide this section in two parts, discussing longitudinally and transversely polarized photons separately.

\subsubsection{Longitudinally polarized photon}

In the presence of angular correlations in the dipole amplitude, we have
\begin{align}
\tilde{F}(\bm{P}, \bm{\Delta}) = \tilde{F}_{0}(P,\Delta) + 2 \tilde{F}_{2} (P,\Delta) \cos 2\theta_{P \Delta} + ...
\end{align}
where
\begin{align}
\frac{\tilde{F}_{0}(P,\Delta)}{(2\pi)^2} =& \int rdr  \ b db \ J_0(Pr) J_0(\Delta b) K_0\left( \varepsilon_f r \right)  D_0(r,b) \,,  \label{ML0} \\
\frac{\tilde{F}_{2}(P,\Delta)}{ (2\pi)^2} =&  \int rdr  \ b db \ J_2(Pr) J_2(\Delta b ) K_0\left( \varepsilon_f r  \right)  D_2(r,b)\,. \label{ML2}
\end{align}
where $D_0(r,b)$ and $D_2(r,b)$ are defined in Eq.\,\eqref{eq:dipoleExpansion}.

This explicitly shows that the coordinate space angular correlations in the dipole amplitude produce angular correlations in momentum space. The expressions for the differential cross section and the elliptic anisotropy for longitudinally polarized photons are then
\begin{align}
\frac{d \sigma_{L,0}}{ d\Omega}  &=  \frac{8 N_c \alpha_{EM}}{(2 \pi)^6} z_0^3 z_1^3 Q^2 Z^2_f  \left( \left | \tilde{F}_{0} \right |^2 + 2 \left| \tilde{F}_{2} \right|^2 \right) \,, \nonumber \\
\frac{d \sigma_{L,2}}{ d\Omega} & =  \frac{8 N_c \alpha_{EM}}{(2 \pi)^6} z_0^3 z_1^3 Q^2 Z^2_f  \Re \left (2 \tilde{F}  _{0}  \tilde{F}^* _{2}\right) \,. \label{CrossLFourier}
\end{align}
The second term in the first line constitutes a small correction to the differential cross section (averaged over angle) due to angular correlations. The second line shows the elliptic anisotropy generated by angular correlations in the dipole amplitude.

\subsubsection{Transversely polarized photon}

To compute the cross section for transversely polarized photons, we follow a similar approach to that of the longitudinal case.
In addition to $\tilde{F}$ we now also need to consider $\tilde{G}$:
\begin{align} \label{GFouriermode}
\tilde{G}(\bm{P}, \bm{\Delta}) = \tilde{G}_{0}(P,\Delta) + 2 \tilde{G}_{2} (P,\Delta) \cos 2\theta_{P \Delta} + ...
\end{align}
where
\begin{align}
\frac{\tilde{G}_{0} (P, \Delta)}{(2\pi)^2} =  \int r dr  \ b db \ J_0(Pr) J_0(\Delta b) \frac{ K_1(\varepsilon_f r) }{r} D_0(r,b)\,, \\
\frac{\tilde{G}_{2}(P, \Delta)}{(2\pi)^2 } = \int r dr  \ b db \ J_2(Pr) J_2(\Delta b) \frac{ K_1(\varepsilon_f r)}{r} D_2(r,b)\,.
\end{align}

To evaluate the expression for the transverse cross section in Eq.\,\eqref{CrossPhotonT}, we need to compute the derivative of $\tilde{G}$ in Eq.\,\eqref{GFouriermode}.

Using the expression for the gradient in polar coordinates: $\partial_{\bm{P}} = \bm{\hat{P}}  \partial_P + \bm{\hat{\theta}}  \frac{1}{P} \partial_\theta $, we find 
\begin{align}
\partial_{\bm{P}} \tilde{G} =&  \left[ \partial_P \tilde{G}_0 + 2 (\partial_P \tilde{G}_2) \cos 2\theta_{P \Delta} \right] \bm{\hat{P}}    \notag\\ 
& - \frac{4\tilde{G}_2}{P} \sin 2 \theta_{P \Delta} \bm{\hat{\theta}} \,.
\end{align}

Keeping terms only up to the second harmonic, we have
\begin{align}
 \left |\partial_{\bm{P}} \tilde{G} \right |^2  = \left| \partial_P \tilde{G}_{0} \right|^2 + 2\left|\partial_P \tilde{G}_{2}\right|^2 + 8 \left|\tilde{G}_{2}\right|^2/P^2 \nonumber \\ +  4 \Re \left( \partial_P \tilde{G}_{0} \partial_P \tilde{G}^*_{2} \right)  \cos 2\theta_{P \Delta} + ...
\end{align}
The derivatives $\partial_P \tilde{G}_0$ and $\partial_P \tilde{G}_{2}$ can be computed using the identities for the derivatives of Bessel functions: $J'_0(z) = - J_1(z)$ and $J'_2(z) = -\frac{1}{2} \left( J_3(z) -J_1(z) \right)$:
\begin{align}
\frac{\partial_P \tilde{G}_0}{(2\pi)^2} = -\int rdr  b db &J_1(Pr) J_0(\Delta b) K_1(\varepsilon_f r) D_0(r,b) \,, \label{PNT0} \\
\frac{\partial_P \tilde{G}_2}{(2\pi)^2} = -\int rdr  b db &\left[ \frac{ J_3(Pr) - J_1(Pr)}{2} \right] J_2(\Delta b) \notag \\
& \hspace{0.5cm} \times K_1(\varepsilon_f r) D_2(r,b) \,.  \label{PNT2}
\end{align}
The components of the differential cross section thus take the following form
\begin{align} \label{CrossTFourier}
 \frac{d \sigma_{T,0}}{ d\Omega}  =& \frac{2 N_c \alpha_{EM}}{(2\pi)^6} Z^2_f z_0 z_1 \nonumber \\   & \times \Bigg\{ \varepsilon_f^2 \zeta^2  \left[\left| \partial_P \tilde{G}_{0}\right|^2 + 2\left|\partial_P \tilde{G}_{2}\right|^2 + 8 \left| \frac{\tilde{G}_{2}}{P}\right|^2\right] \nonumber \\ &  ~~~~~~ + m_f^2 \left( \left|\tilde{F}_{0}\right|^2 + 2 \left|\tilde{F}_{2}\right|^2 \right) \Bigg\} \,,\nonumber \\
 \frac{d \sigma_{T,2}}{ d\Omega}  =&  \frac{2 N_c \alpha_{EM}}{(2\pi)^6}   Z^2_f  z_0 z_1  \nonumber \\   & \times \bigg\{  \varepsilon_f^2 \zeta^2 \Re \left( 2\partial_P \tilde{G}_{0} \partial_P \tilde{G}^*_{2} \right) \notag\\
    & ~~~~~~ + m_f^2  \Re \left( 2\tilde{F}  _{0} \tilde{F}^*  _{2}  \right) \bigg\}\,.
\end{align}

As in the case of longitudinally polarized photons, the additional terms in the (angle averaged) differential cross section provide a small correction. Again, the elliptic anisotropy arises from angular correlations in the dipole amplitude.

To proceed further, one needs an explicit form of the dipole amplitude. In the next section we introduce two specific models. We will then investigate analytic properties of above cross sections in Section \ref{sec:analytics}.  In Section \ref{sec:numrel} we will evaluate the cross sections numerically and present detailed results as functions of $P$ and $\Delta$.

\section{Review of Dipole Models}
\label{sec:dipoleModels}

The analytic properties of the differential cross sections and elliptic anisotropies in Eqs.\,(\ref{CrossLFourier}) and (\ref{CrossTFourier}) depend on the light-cone wave function and the dipole amplitude, which encode information on the projectile and target. Because in this work we focus on understanding the analytic structure of the diffractive dijet cross sections, we will not perform complex numerical calculations, as done e.g. in \cite{Mantysaari:2019csc}, but introduce relatively simple models for the dipole amplitude. We focus on dipole amplitudes of the form
\begin{align}
D(\bm{r},\bm{b}) = 1-e^{-\mathcal{N}_0(r,b)-\mathcal{N}_2(r,b) \cos 2\theta_{rb} }\,.\label{DipoleForm}
\end{align}
This parametrization is appropriate for small dipole sizes compared to the color charge density gradient of the target. The second term in the exponent contains the angular correlations between $\bm{r}$ and $\bm{b}$, which will ultimately produce the angular correlations in the cross sections.

The dipole in Eq.\,(\ref{DipoleForm}) admits a simple form for the modes $D_0$ and $D_2$, introduced in Eq.\,\eqref{eq:dipoleExpansion}. By proper projection and using the integral representations of modified Bessel functions of the first kind, $I_0(z)$ and $I_1(z)$ (Eq.\,(\ref{BesselI})), we find
\begin{align} \label{Dprojections}
D_0(r,b) = 1-e^{-\mathcal{N}_0(r,b)}I_0(\mathcal{N}_2(r,b)) \,,\nonumber \\
D_2(r,b) = e^{-\mathcal{N}_0(r,b)}I_1(\mathcal{N}_2(r,b))\,.
\end{align}
In the limit of small dipole sizes we have
\begin{align}
\label{ExpandedDProjection}
    D_0(r,b) \approx \mathcal{N}_0(r,b)\,, \nonumber \\
    D_2(r,b) \approx \frac{1}{2}\mathcal{N}_2(r,b)\,.
\end{align}

In the following we introduce explicit forms for $\mathcal{N}_0$ and $\mathcal{N}_2$ based on the Golec-Biernat Wusthoff and impact parameter dependent McLerran Venugopalan model. 

\subsection*{Golec-Biernat Wusthoff model}

A very simple model to describe the dipole amplitude is the Golec-Biernat Wusthoff (GBW) model \cite{GolecBiernat:1998js,GolecBiernat:1999qd}, where, after introducing an impact parameter dependence, $\mathcal{N}_0$ in Eq.\,\eqref{DipoleForm} takes the phenomenologically motivated form 
\begin{align} \label{DipoleGBWnonangular}
\mathcal{N}_0(r,b)=\frac{1}{4}Q_s^2 r^2 T(b)\,,
\end{align}
with $Q_s$ the saturation scale at zero impact parameter, and $T(b)$ the transverse spatial profile of the target, which we assume to be isotropic.

The GBW model does not contain angular correlations between impact parameter and dipole orientation. One could add the angular correlations by hand as was done in \cite{Altinoluk:2015dpi} by choosing for example
\begin{align}\label{eq:GBWN2}
\mathcal{N}_2(r,b)= \frac{c}{4}Q_s^2 r^2 T(b)\,,
\end{align}
where $c$ characterizes the strength of angular correlations ($-1<c<1$).

\subsection*{Impact parameter dependent McLerran Venugopalan model}

In \cite{Iancu:2017fzn} the authors computed the dipole amplitude for an impact parameter dependent McLerran Venugopalan (IP-MV) model.

They found a dipole amplitude of the form of Eq.\,\eqref{DipoleForm} with
\begin{align}
\mathcal{N}_0(r,b) &= \frac{1}{4}Q_s^2 r^2 \log\left( \frac{1}{r^2 m^2} + e \right) T(b) + ... \label{IPMVnoangular} \\
\mathcal{N}_2(r,b) &= \frac{1}{4} Q_s^2 r^2 \frac{1}{6 m^2} \left[ \frac{d^2}{db^2} - \frac{1}{b}\frac{d}{db} \right] T(b) + ... \label{IPMVangular}
\end{align}
where $m$ is an infrared regulator, and the factor of $e$ in the logarithm is included to regulate the divergence for dipoles of sizes larger than $1/m$. We provide details on the derivation of \eqref{IPMVnoangular} and \eqref{IPMVangular} in Appendix \ref{app:ipmv}.

The ellipses in Eqs.\,\eqref{IPMVnoangular} and \eqref{IPMVangular} denote corrections which can be ignored if the gradients are slowly varying with respect to the confinement scale $1/m$. The logarithmic factor in $\mathcal{N}_0$ arises from the non-local interactions of the dipole with the color charges. (A non-local operation is involved in determining the Wilson lines from the color charges. See Appendix \ref{app:ipmv}.) 

It is important to point out that the angular correlations $\mathcal{N}_2$ are proportional to gradients of the color charge density. This is because the dipole will probe regions of different color charge density depending on its orientation relative to the impact parameter vector. The larger the variation of the color charge density around $\mathbf{b}$, the larger the angular modulations encoded in $\mathcal{N}_2$. Furthermore, the angular correlations are suppressed for large confining scale $m$. As $m \rightarrow 0$, $\mathcal{N}_2$ does not diverge, but $m$ is replaced as regulator by the finite system size, or $1/R$ (see details in Appendix \ref{app:ipmv}).

\section{Analytic properties}
\label{sec:analytics}

Recently, coherent diffractive dijet production has been studied numerically. Some interesting properties of the differential cross section and elliptic anisotropy were observed \cite{Mantysaari:2019csc}. Our goal is to analyze the properties of the dijet production cross sections and their angular dependence based on their analytic structure. In particular we will investigate the dependence on the saturation scale $Q_s$, as well as the photon virtuality $Q^2$, photon polarization, (anti-)quark mass $m_f$, and the geometry of the target's color charge density. 

We will begin our analysis with the GBW model for which the analytic calculations are simpler, and the results will reveal interesting properties of the differential cross section and elliptic anisotropy. We then move to the more complex IP-MV model in order to study a more realistic scenario.

\subsection{Golec-Biernat Wusthoff model}

The simplest model we consider is the GBW model. Even for this simple case the remaining integrals in Eqs.\,\eqref{ML0},\eqref{ML2},\eqref{PNT0}, and \eqref{PNT2} cannot be solved analytically in general. However, in the limit where the saturation scale $Q_s$ is much smaller than the photon virtuality or quark mass, analytic expressions can be found. We present results in this limit in the next section and discuss the consequences of relaxing them thereafter.

\subsubsection{Large photon virtuality or massive quarks: $Q_s \ll \varepsilon_f $}
\label{sec:largevirtuality}

If $Q_s \ll \varepsilon_f$ the Bessel functions $K_0$ and $K_1$ suppress contributions from $r\gtrsim 1/Q_s$ in the $r$-integrals in Eqs.\,\eqref{ML0},\eqref{ML2},\eqref{PNT0}, and \eqref{PNT2}. For $r \lesssim 1/\varepsilon_f \ll 1/Q_s$, one can expand the dipole to quadratic order (Eq.\,\eqref{ExpandedDProjection}) and obtain
\begin{align}
D_0(r,b)  \approx \frac{1}{4} Q^2_s r^2 T(b)\,, \nonumber \\
D_2(r,b)  \approx \frac{c}{8} Q^2_s r^2 T(b)\,. 
\label{SmallDipole}
\end{align}
In this limit the dependence of the dipole amplitude on $r$ and $b$ factorizes, and one finds the following expressions for the differential cross sections (we provide further details on the calculation of the following results in Appendix \ref{app:analyticsupplement})
\begin{align}
\frac{d \sigma_{L,0} }{d \Omega}=&\frac{8 N_c \alpha_{EM}}{(2\pi)^4} Z^2_f Q^2 z_0^3 z_1^3 Q_s^4 \frac{ \left( P^2-\varepsilon^2_f \right)^2}{\left(P^2+\varepsilon^2_f \right)^6}  |\tilde{T}(\Delta)|^2 \,,\\
\frac{d \sigma_{T,0} }{d \Omega} =  &\frac{2 N_c \alpha_{EM}}{(2\pi)^4} Z^2_f  z_0 z_1   Q_s^4  \nonumber \\ &\times\left\lbrace  \frac{4 \varepsilon_f^2 \zeta^2 P^2 \varepsilon^2_f + m_f^2 \left( \varepsilon^2_f-P^2 \right)^2}{\left(P^2+\varepsilon^2_f \right)^6}  \right\rbrace |\tilde{T}(\Delta)|^2 \label{ToyGBWsigma0}\,,
\end{align}
and the elliptic anisotropies
\begin{align}
\frac{d \sigma_{L,2}  }{d \Omega}=&- \frac{8 N_c \alpha_{EM}}{(2 \pi)^4} z_0^3 z_1^3 Q^2 Z^2_f Q_s^4  \nonumber \\ & \times 2c \frac{( P^2- \varepsilon_f^2)P^2}{(P^2+\varepsilon^2_f)^6} \tilde{T}(\Delta)  \tilde{T}_2(\Delta) \label{ToyGBWsigma2L}\,,\\
\frac{d \sigma_{T,2}}{ d\Omega}  = &\frac{2 N_c \alpha_{EM}}{(2\pi)^4} Z^2_f  z_0 z_1 (\varepsilon^2_f \zeta^2-m_f^2)  Q_s^4 \nonumber \\  &\times 2c  \frac{( P^2- \varepsilon_f^2)P^2}{(P^2+\varepsilon^2_f)^6} \tilde{T}(\Delta)  \tilde{T}_2(\Delta) \label{ToyGBWsigma2T}\,,
\end{align}
where $\tilde{T}_2(\Delta)$ is the 2nd order Hankel transform of $T(b)$, i.e., $\tilde{T}_2(\Delta) = 2\pi \int bdb J_2(\Delta b) T(b)$. \\

As expected from the factorization in $b$ and $r$, in this limit the $P$ and $\Delta$ dependencies factorize, and both the differential cross section and elliptic anisotropy grow as $Q_s^4$. This growth will eventually be tamed by saturation effects as we discuss in the next subsection.

The $P$-dependence probes the projectile - it is sensitive to the photon's polarization and virtuality, as well as the quark mass encoded in $\varepsilon_f$. The $P$-dependence of the differential cross section for longitudinally polarized photons develops a dip at $P=\varepsilon_f$. The location of the dip depends directly on the virtuality $Q^2$ and the mass $m_f$ of the quark.
This feature is absent in the transversely polarized photon case, which has two contributions, only one of which exhibits a dip. The elliptic anisotropies for both polarizations change sign at $P=\varepsilon_f$, a feature also observed in the IPSat model in \cite{Mantysaari:2019csc}.  The sign of the elliptic anisotropy in the transverse case depends on the ratio $\varepsilon_f \zeta / m_f$.

At large $P$ the differential cross section and elliptic anisotropy decay with the power law $1/P^8$, while for small $P$ the former is constant, and the latter vanishes.
Similarly, at $Q^2 \gg P^2$, both cross sections scale as $Q^{-6}$, while for smaller $Q^2 \ll P^2$, they scale as $Q^2$. Both elliptic anisotropies scale as $Q^{-8}$ for $Q^2 \gg P^2$ and $Q^2$ for $Q^2 \ll P^2$ in the GBW model.
We note that detailed scaling relations with $Q^2$ and the mass number $A$ for the related process of diffractive vector meson production were discussed in \cite{Mantysaari:2017slo}. 
Information on the target is encoded in the $\Delta$-dependence of the differential cross section, which involves the Fourier transform of the color charge density profile. The $\Delta$-dependence of the elliptic anisotropy provides access to target properties via the product $\tilde{T}(\Delta) \tilde{T}_2(\Delta)$. This particular dependence on the target geometry is a consequence of how we introduced the angular dependence in the GBW model in Eq.\,\eqref{eq:GBWN2}. In the more realistic IP-MV model discussed below, the elliptic anisotropy will only depend on $\tilde{T}(\Delta)$.
 
\subsubsection{Approaching the saturated regime: $Q_s \sim \varepsilon_f$}

The case where the saturation scale is of the same order as $\varepsilon_f$ is more difficult to study analytically, as one cannot expand the dipole amplitude to quadratic order, because dipoles of size $r\sim 1/Q_s$ will contribute to the cross sections. One should use the full expressions in Eq.\, \eqref{Dprojections} in which the $r$ and $b$ dependencies do not factorize. Therefore,
the $P$-dependence and $\Delta$-dependence of the differential cross section and elliptic anisotropy no longer factorize either.

For fixed dipole size $r$, the dipole amplitude no longer grows as $Q^2_s$, but it saturates to unity, thus slowing down the growth of the differential cross section and elliptic anisotropies as the saturation scale $Q_s$ is increased. 

While in the high virtuality or large quark mass limit the dominant momentum scale is $\varepsilon_f$, we now have a competition between the two scales $Q_s$ and $\varepsilon_f$. This will be reflected in the $Q_s$-dependence of observables. For instance, the dip in $P$ for the differential cross section in the longitudinally polarized case and the change in sign in the elliptic anisotropy will also be sensitive to the saturation scale $Q_s$. Their locations will shift to larger values of $P$ as the saturation scales increases. This can also be justified mathematically by observing that the location of the maximum of the product of dipole amplitude and light-cone wave function for longitudinally polarized photons,
\begin{align}
    K_0(r) (1-e^{-\frac{1}{4}r^2Q^2_s T(b)})\,,
\end{align}
will shift towards smaller values of $r$ as $Q_s$ increases. Thus, the Fourier transform of this product will have a zero at a larger value of $P$, causing the shift of the dip.

Another interesting feature is that the dipole amplitude is no longer proportional to the target color charge profile $T(b)$, thus producing a more complex $\Delta$-dependence of the differential cross section and elliptic anisotropy. The $\Delta$-dependence will not only depend on the geometry of the profile but also on saturation effects. 

\subsection{Impact parameter dependent McLerran Venugopalan model}
\label{sec:IPMVcross}

In this section we consider an approximation based on a parametric estimate of the effect of the logarithm in the IP-MV model, which distinguishes the outcome of the model from the result in the previously discussed GBW model. To gain more insight into the features of this model, a numerical study is necessary, which will be presented in the next section.

\subsubsection{Large photon virtuality or massive quarks: $Q_s \ll \varepsilon_f $,} \label{sec:IPMVanalytic}

As done in the GBW model above, we expand the dipole for small sizes $r$ and arrive at
\begin{align}
D_0(r,b) \approx & \frac{1}{4} Q^2_s r^2 \log\left(\frac{1}{m^2 r^2} + e \right)\,, \nonumber \\
D_2(b,r) \approx &\frac{1}{8} Q_s^2 r^2 \frac{1}{6 m^2} \left[ \frac{d^2}{db^2} - \frac{1}{b}\frac{d}{db} \right] T(b)\,. \label{SmallDipoleIPMV}
\end{align}
The presence of the logarithmic factor in $D_0$ makes the analytic evaluation of $\tilde{F}_0$ \eqref{ML0} and $\partial_P \tilde{G}_0$ \eqref{PNT0} difficult. In Appendix \ref{app:analyticsupplement} we argue that the effect of the logarithmic factor is to enhance the value of $\tilde{F}_0$ and $\partial_P \tilde{G}_0$ and to shift the zero of $\tilde{F}_0$ to a larger value of $P$ (See Eqs.\,\eqref{F0IPMVmodel} and \eqref{G0IPMVmodel}). We arrive at approximate expressions for the differential cross sections: 
\begin{align}
\frac{d \sigma_{L,0} }{d \Omega} \approx &\frac{8 N_c \alpha_{EM}}{(2\pi)^4} Z^2_f Q^2 z_0^3 z_1^3 Q_s^4  |\tilde{T}(\Delta)|^2 \nonumber \\ &  \times \frac{ C^2_1\left( P^2-\xi^2 \varepsilon^2_f \right)^2}{\left(P^2+\xi^2 \varepsilon^2_f \right)^6} \label{ToyIPMVsigma0L}  \,,\\
\frac{d \sigma_{T,0} }{d \Omega} \approx  &\frac{2 N_c \alpha_{EM}}{(2\pi)^4} Z^2_f  z_0 z_1   Q_s^4  \varepsilon^2_f |\tilde{T}(\Delta)|^2  \nonumber \\ & \times \frac{4 C^2_2 \varepsilon_f^2 \zeta^2 P^2 \xi^2 \varepsilon^2_f + C^2_1 m_f^2 \left( \xi^2 \varepsilon^2_f-P^2 \right)^2}{\left(P^2+\xi^2 \varepsilon^2_f \right)^6}  \label{ToyIPMVsigma0T}\,,
\end{align}
where $\xi$ characterizes the shift, and $C_1$ and $C_2$ are the enhancement factors. In Appendix \ref{app:analyticsupplement}, we show that $\xi > 1$ and $C_2 > C_1 > 1$.

Similarly, one finds that the elliptic anisotropies are given by
\begin{align}
\frac{d \sigma_{L,2} }{d \Omega} \approx& - \frac{8 N_c \alpha_{EM}}{(2 \pi)^4} z_0^3 z_1^3 Q^2 Z^2_f Q_s^4  \frac{\Delta^2 |\tilde{T}(\Delta)|^2}{3m^2}  \nonumber \\ &  \times \frac{C_1 ( P^2-\xi^2 \varepsilon_f^2)P^2}{(P^2+\varepsilon^2_f)^3 (P^2+\xi^2 \varepsilon^2_f)^3} \label{ToyIPMVsigma2L}\,,\\
\frac{d \sigma_{T,2}}{ d\Omega}  \approx& \frac{2 N_c \alpha_{EM}}{(2\pi)^4} Z^2_f  z_0 z_1 Q_s^4 \frac{\Delta^2 |\tilde{T}(\Delta)|^2}{3m^2} P^2 \nonumber \\  & \times \bigg[C_2\zeta^2\varepsilon^2_f \xi (P^2-\varepsilon^2_f) -C_1 m_f^2(P^2-\xi^2\varepsilon^2_f)\bigg]\notag\\ &\times\bigg[(P^2+\varepsilon^2_f)(P^2+\xi^2 \varepsilon^2_f)\bigg]^{-3} \label{ToyIPMVsigma2T}\,.
\end{align}
As in the GBW model for $Q_s\ll\varepsilon_f$, the $P$ and $\Delta$ dependencies factorize. The $P$-dependence is sensitive to the light-cone wave function and the logarithm in the dipole amplitude, as manifest in the dependence on $\xi$. In the longitudinal case, the differential cross section displays a dip at $P=\xi \varepsilon_f$ (greater value of $P$ compared to GBW - a similar behavior was observed in the location of the dip in the IPSat vs CGC results in \cite{Mantysaari:2019csc}). As in the GBW model, the transverse cross section does not display a dip. 

An interesting feature of the IP-MV model in this limit is that the elliptic anisotropy is highly sensitive to the infrared regulator $m$. One should mention that these formulas break down when $m \ll \Delta$, since Eq.\,\eqref{IPMVangular} has been obtained assuming small momentum transfer kicks to the dipole (See also last paragraph of Appendix \ref{app:ipmv}). 

The elliptic anisotropy for the longitudinal photon changes sign at $P=\xi \varepsilon_f$ (greater value of $P$ compared to GBW, the shift was also seen in \cite{Mantysaari:2019csc} from IPSat to CGC.).

The behavior of the elliptic anisotropy for the transverse polarization is more subtle. Let us assume that $\xi \sim 1.7$ and $C_2/C_1 \sim 1.2$ (see Appendix \ref{app:analyticsupplement}). At small $P$, we have
\begin{align}
\frac{d \sigma_{T,2}}{d \Omega}(P\ll \varepsilon_f) >0 \Leftrightarrow (\zeta^2 \varepsilon^2_f )/m^2_f <  \xi C_1/C_2 \sim 1.4 \,. \label{Cond1}
\end{align}
This implies that at low $Q^2$, the elliptic anisotropy is positive for small $P$. In contrast, for sufficiently large $Q^2$, the anisotropy at small $P$ is negative. 

At large $P$ we have
\begin{align}
    \frac{d \sigma_{T,2}}{d \Omega}(P\gg \varepsilon_f) >0 \Leftrightarrow m^2_f/(\zeta^2 \varepsilon^2_f ) <  \xi C_2/C_1 \sim 2 \,.
    \label{Cond2}
\end{align}
For our choice of $\xi$ and $C_2/C_1$, this condition is always satisfied, and thus one expects the elliptic anisotropy to remain positive. Therefore, for transverse polarization we expect a change in sign in the elliptic anisotropy from negative to positive for high virtuality $Q^2$. For low $Q^2$, we expect the elliptic anisotropy to be positive for all $P$. Such behavior has been observed in the recent numerical study in \cite{Mantysaari:2019csc}.
We observe the same scaling with $Q^2$ and $P$ for the differential cross section and elliptic anisotropy as in the GBW model. 

The $\Delta$-dependence of the differential cross section is similar to that of the GBW case; the Fourier transform $\tilde{T}(\Delta)$ of the color charge density appears and introduces the sensitivity to the details of the geometry of the target's color charge density. The elliptic anisotropy differs from that of GBW, in which the angular correlations where included by hand. In the IP-MV case, where the angular correlations emerge as a result of the finite size gradients of the profile, we only find a dependence on $\tilde{T}(\Delta)$, not on $\tilde{T}_2(\Delta)$. Eqs.\,\eqref{ToyIPMVsigma2L} and \eqref{ToyIPMVsigma2T} also show that as the momentum transfer $\Delta$ goes to zero (exact correlation limit), the elliptic anisotropy vanishes. The position of the maximum as a function of $\Delta$ will depend on the details of the profile.

\begin{figure*}[!htb]
    \centering
    \begin{minipage}[t]{0.45\textwidth}
        \centering
        \includegraphics[width=8cm]{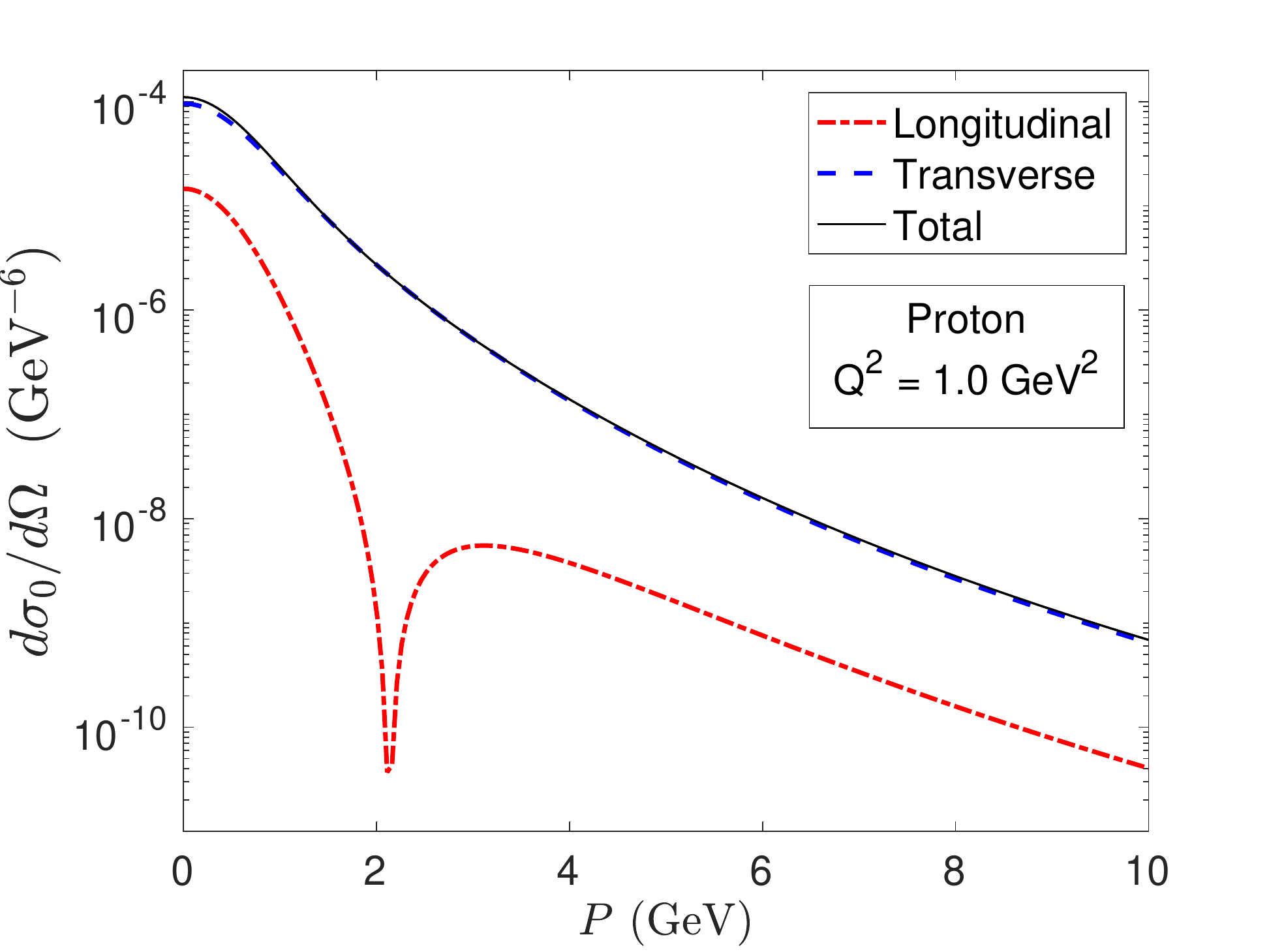}
        \caption{Differential cross section $d \sigma_0 / d \Omega$ as a function of $P$ for a proton target. Here $\Delta = 0.1$ GeV and $Q^2= 1.0$ GeV$^2$.}  \label{sigma_0_pdep_proton_Q1}
        \includegraphics[width=8cm]{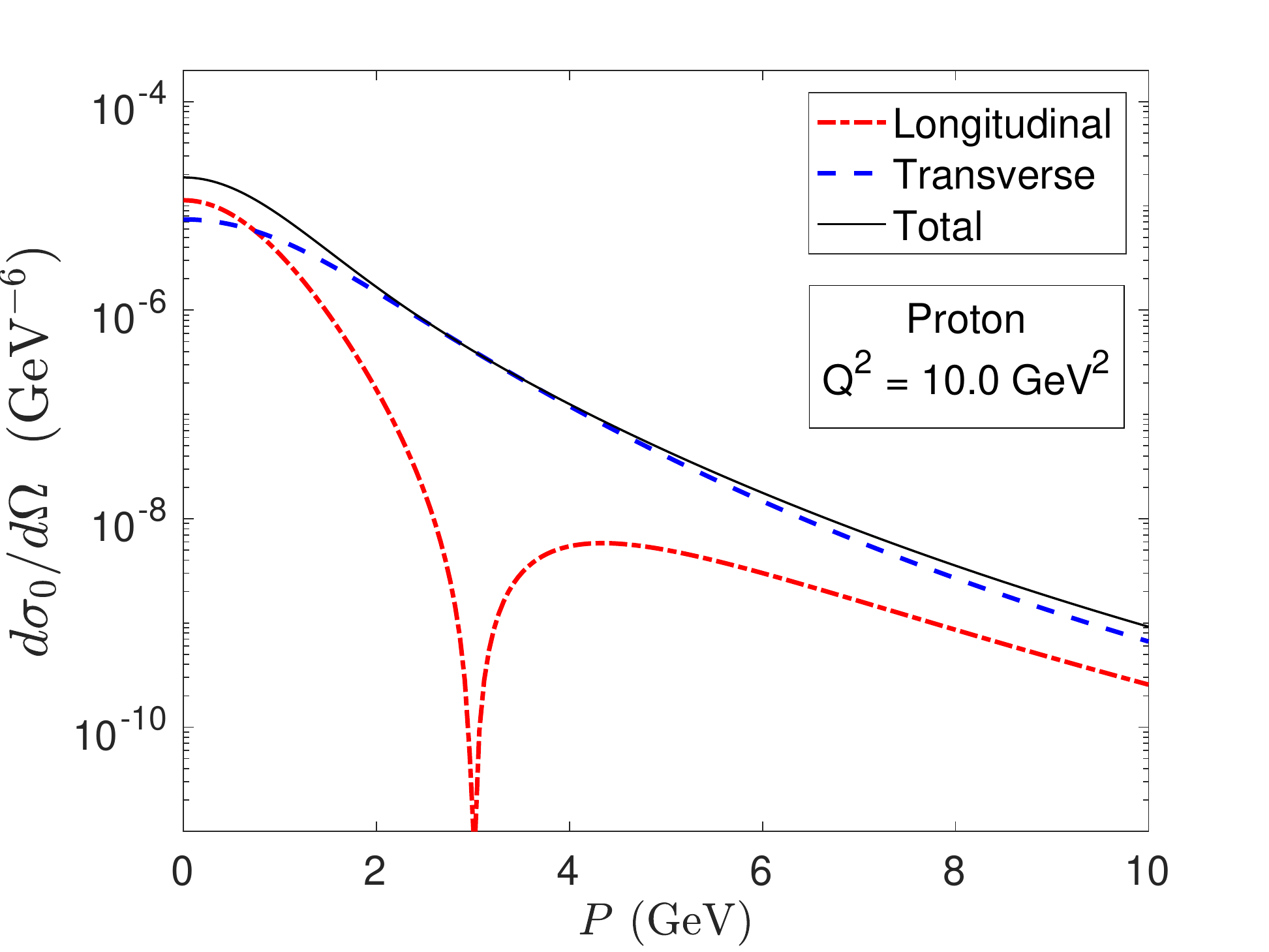}
        \caption{Differential cross section $d \sigma_0 / d \Omega$ as a function of $P$ for a proton target. Here $\Delta = 0.1$ GeV and $Q^2= 10.0$ GeV$^2$.}  \label{sigma_0_pdep_proton_Q10}
    \end{minipage}
    \hspace{0.5cm}
    \begin{minipage}[t]{0.45\textwidth}
        \centering
        \includegraphics[width=8cm]{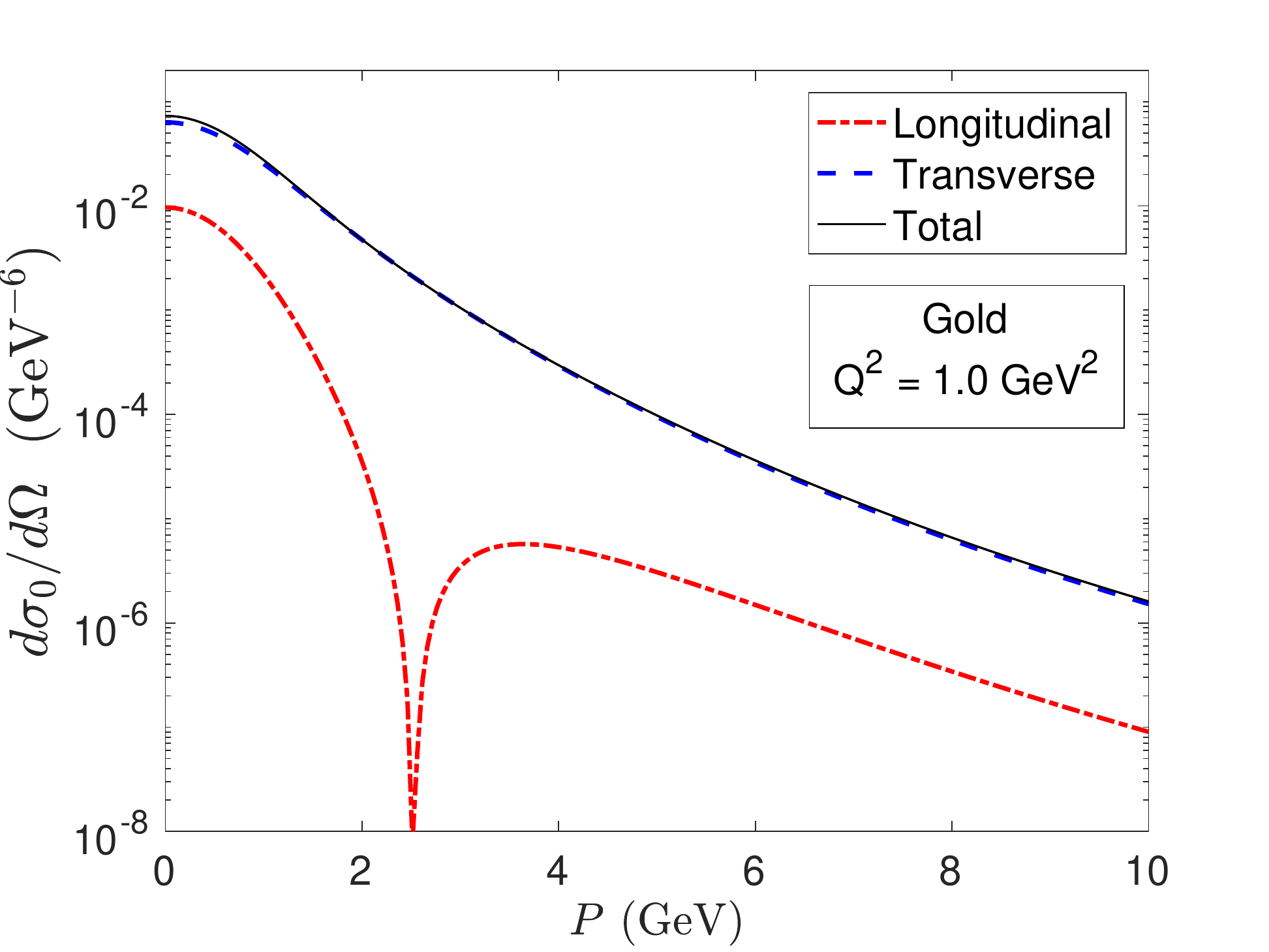}
        \caption{Differential cross section $d \sigma_0 / d \Omega$ as a function of $P$ for a gold target. Here $\Delta = 0.1$ GeV and $Q^2= 1.0$ GeV$^2$.}  \label{sigma_0_pdep_Au_Q1}
        \includegraphics[width=8cm]{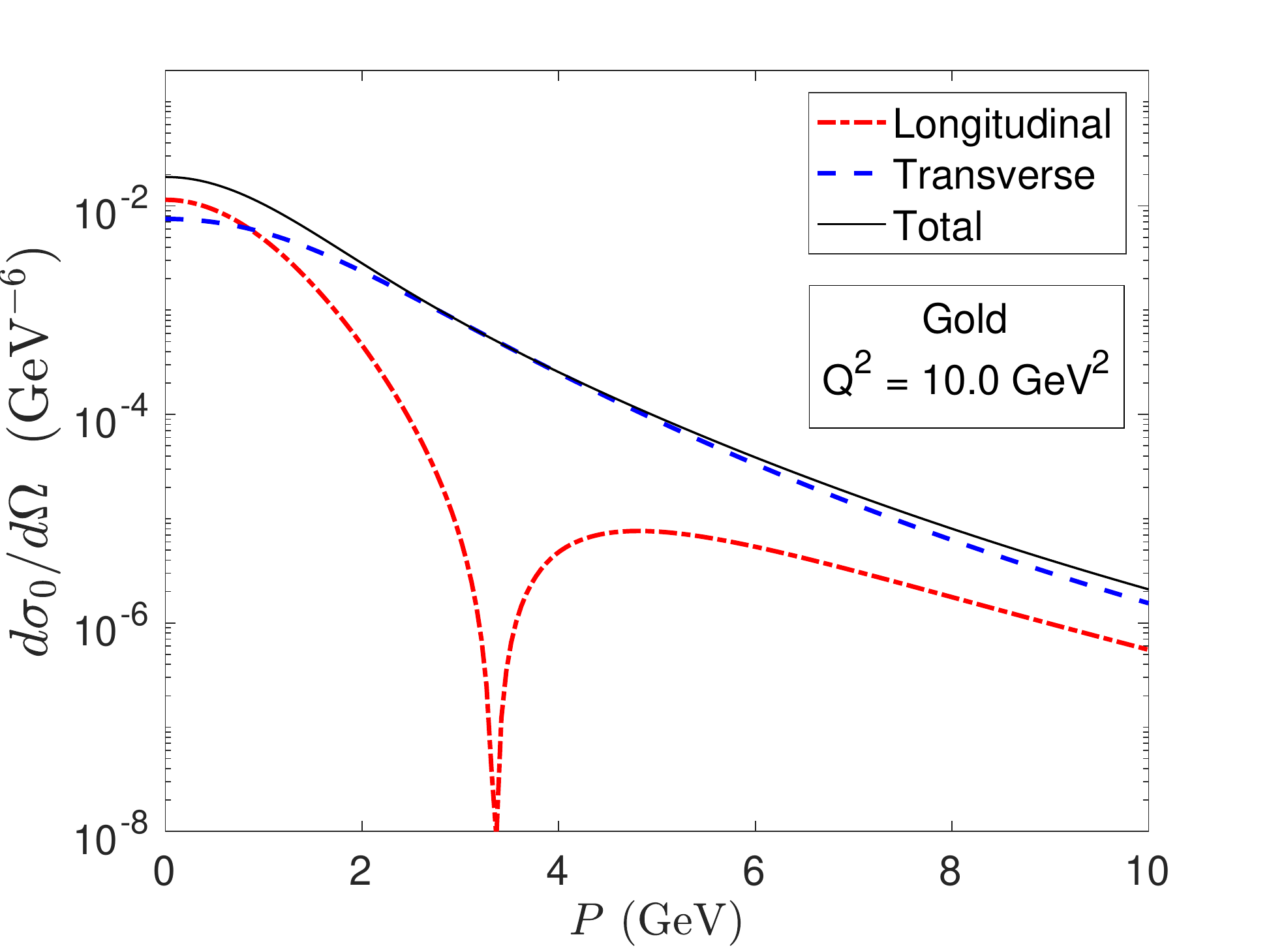}
        \caption{Differential cross section $d \sigma_0 / d \Omega$ as a function of $P$ for a gold target. Here $\Delta = 0.1$ GeV and $Q^2= 10.0$ GeV$^2$.} \label{sigma_0_pdep_Au_Q10}
    \end{minipage}
\end{figure*}

\begin{figure*}[!htb]
    \centering
    \begin{minipage}[t]{0.45\textwidth}
        \centering
        \includegraphics[width=8cm]{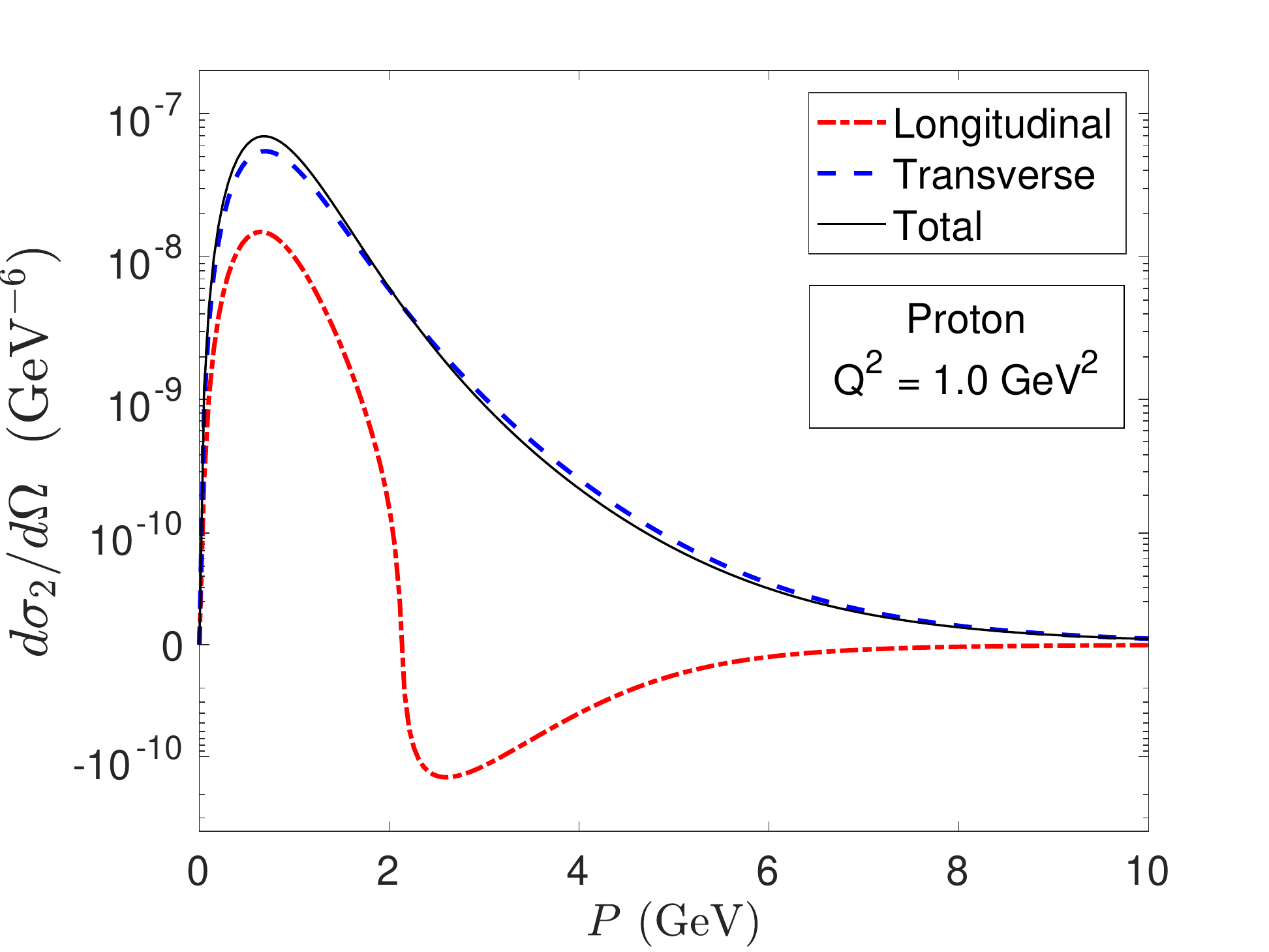}
        \caption{Elliptic anisotropy $d \sigma_2 / d\Omega$ as a function of P for a proton target. Here $\Delta = 0.1$ GeV and $Q^2= 1.0$ GeV$^2$.} \label{sigma_2_pdep_proton_Q1}
        \includegraphics[width=8cm]{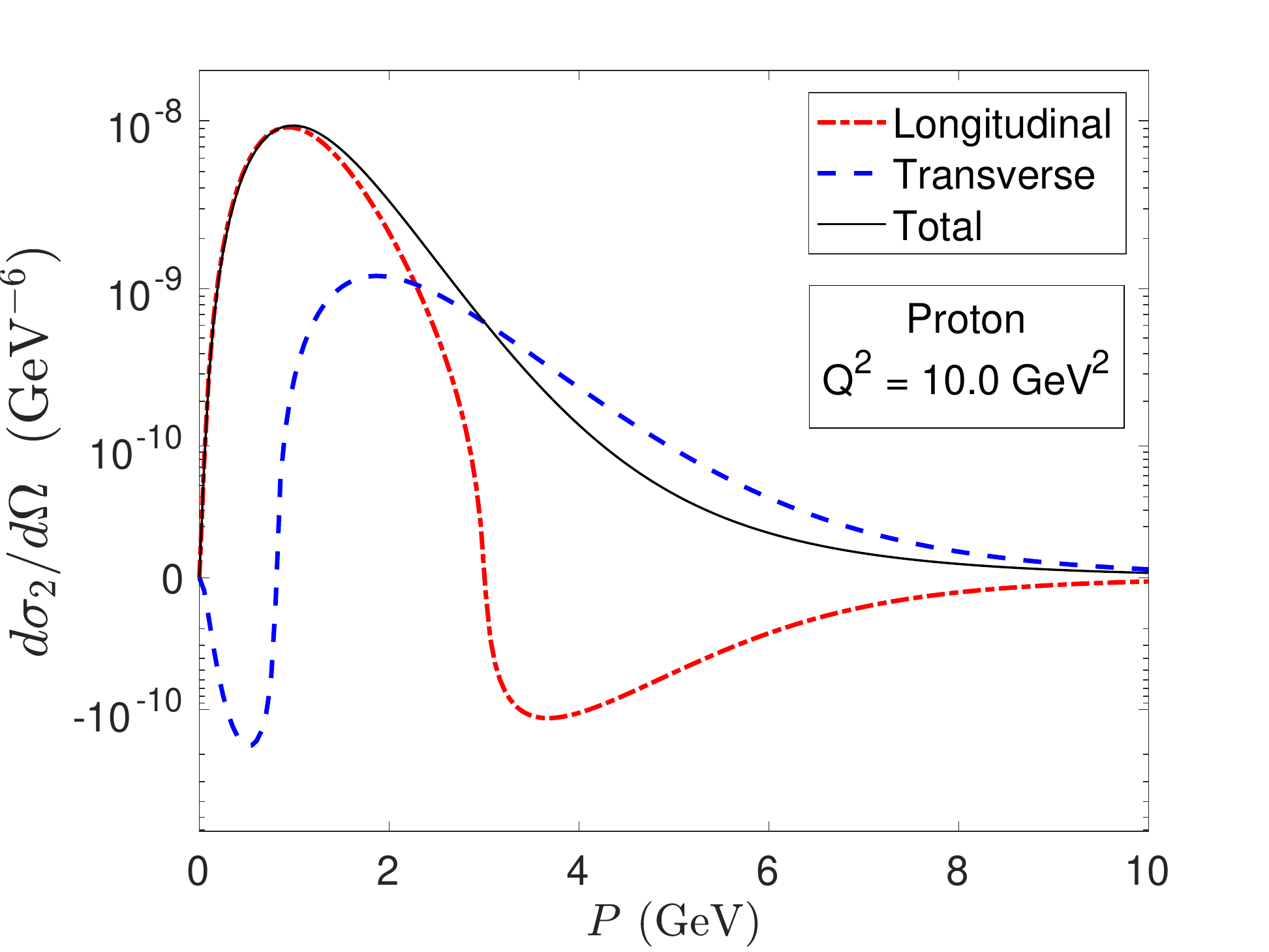}
        \caption{Elliptic anisotropy $d \sigma_2 / d\Omega$ as a function of P for a proton target. Here $\Delta = 0.1$ GeV and $Q^2= 10.0$ GeV$^2$.} \label{sigma_2_pdep_proton_Q10}
    \end{minipage}
    \hspace{0.5cm}
    \begin{minipage}[t]{0.45\textwidth}
        \centering
        \includegraphics[width=8cm]{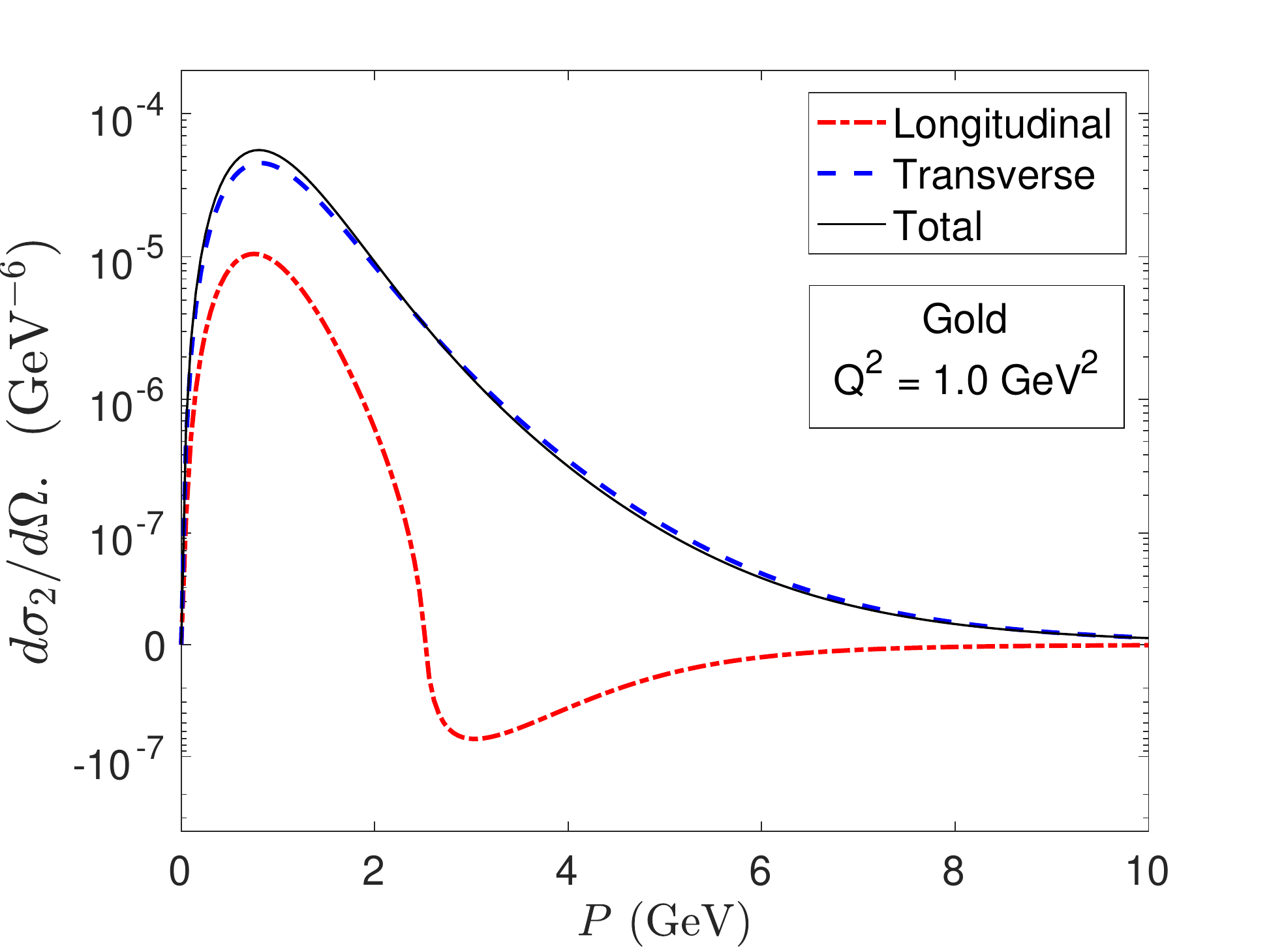}
        \caption{Elliptic anisotropy $d \sigma_2 / d\Omega$ as a function of P for a gold target.   Here $\Delta = 0.1$ GeV and $Q^2= 1.0$ GeV$^2$.}  \label{sigma_2_pdep_Au_Q1}
        \includegraphics[width=8cm]{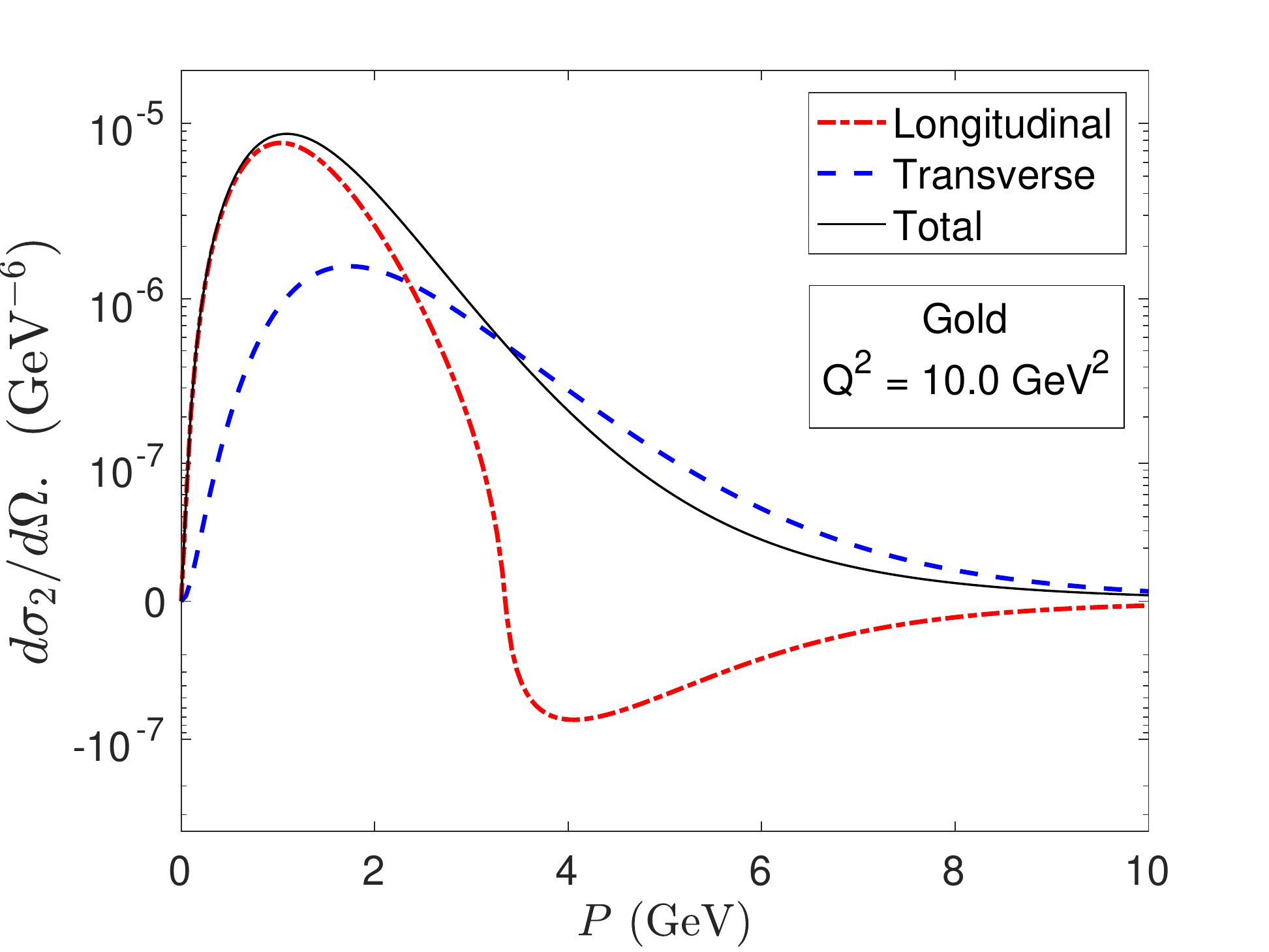}
        \caption{Elliptic anisotropy $d \sigma_2 / d\Omega$ as a function of P for a gold target. Here $\Delta = 0.1$ GeV and $Q^2= 10.0$ GeV$^2$.} \label{sigma_2_pdep_Au_Q10}
    \end{minipage}
\end{figure*}

\begin{figure*}[!htb]
    \centering
    \begin{minipage}[t]{0.45\textwidth}
        \centering
         \includegraphics[width=8cm]{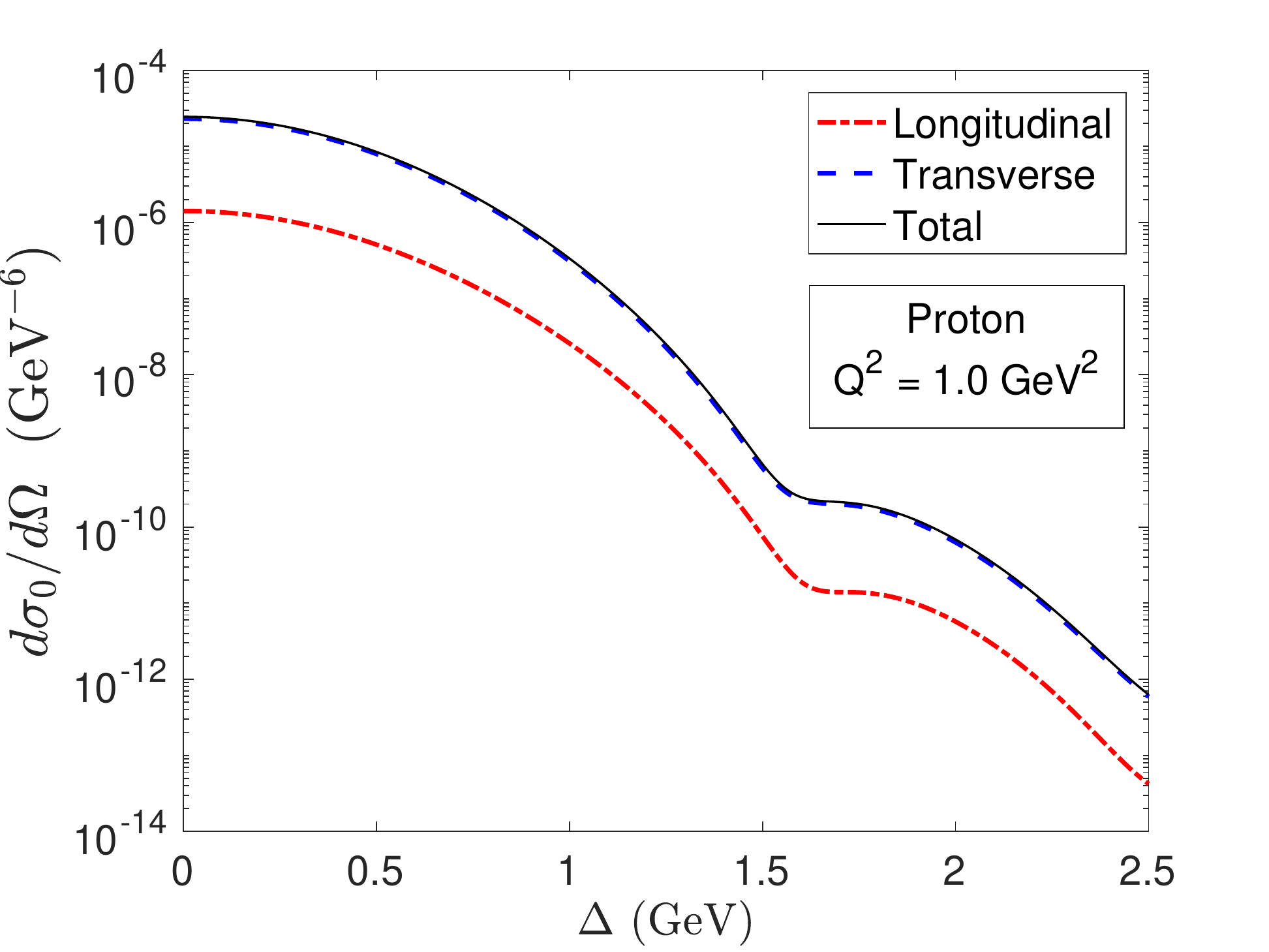}
        \caption{Differential cross section $d \sigma_0 / d \Omega$ as a function of $\Delta$ for a proton target. Here $P = 1.0$ GeV and $Q^2= 1.0$ GeV$^2$.}  \label{sigma_0_Deltadep_proton_Q1}
        \includegraphics[width=8cm]{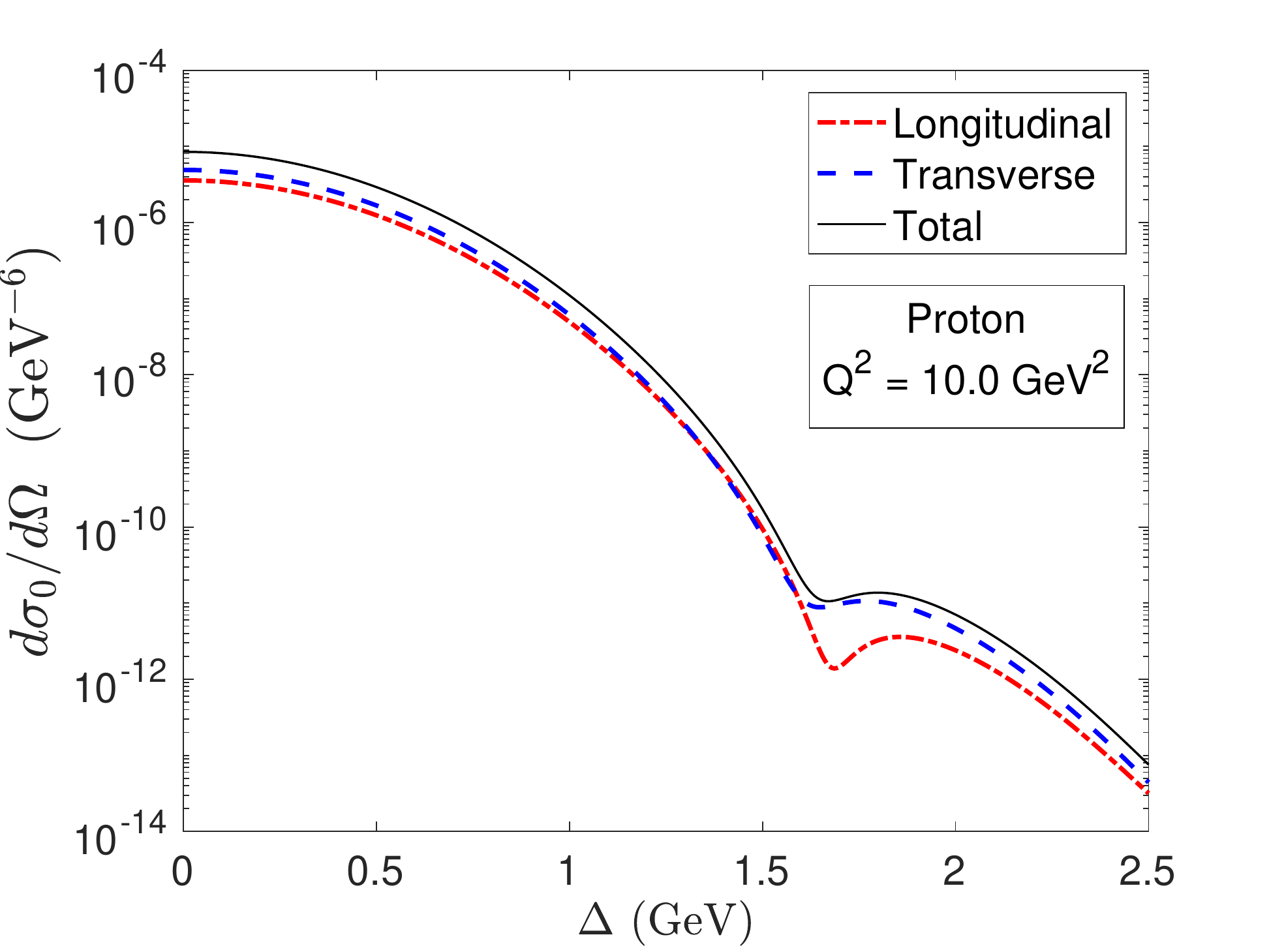}
        \caption{Differential cross section $d \sigma_0 / d \Omega$ as a function of $\Delta$ for a proton target. Here $P = 1.0$ GeV and $Q^2= 10.0$ GeV$^2$.}  \label{sigma_0_Deltadep_proton_Q10}
    \end{minipage}
    \hspace{0.5cm}
    \begin{minipage}[t]{0.45\textwidth}
        \centering
        \includegraphics[width=8cm]{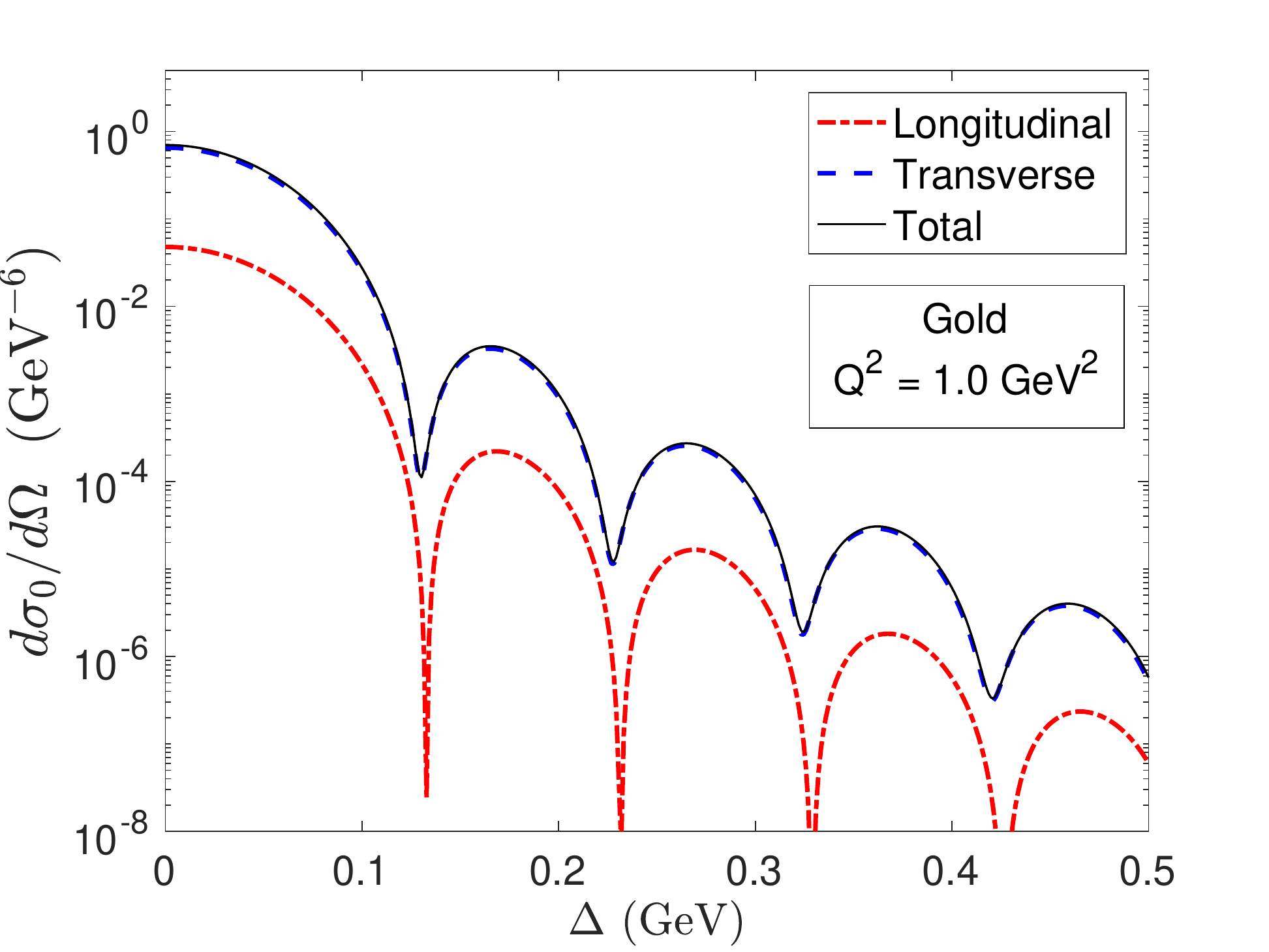}
        \caption{Differential cross section $d \sigma_0 / d \Omega$ as a function of $\Delta$ for a gold target. Here $P = 1.0$ GeV and $Q^2= 1.0$ GeV$^2$.}  \label{sigma_0_Deltadep_Au_Q1}
        \includegraphics[width=8cm]{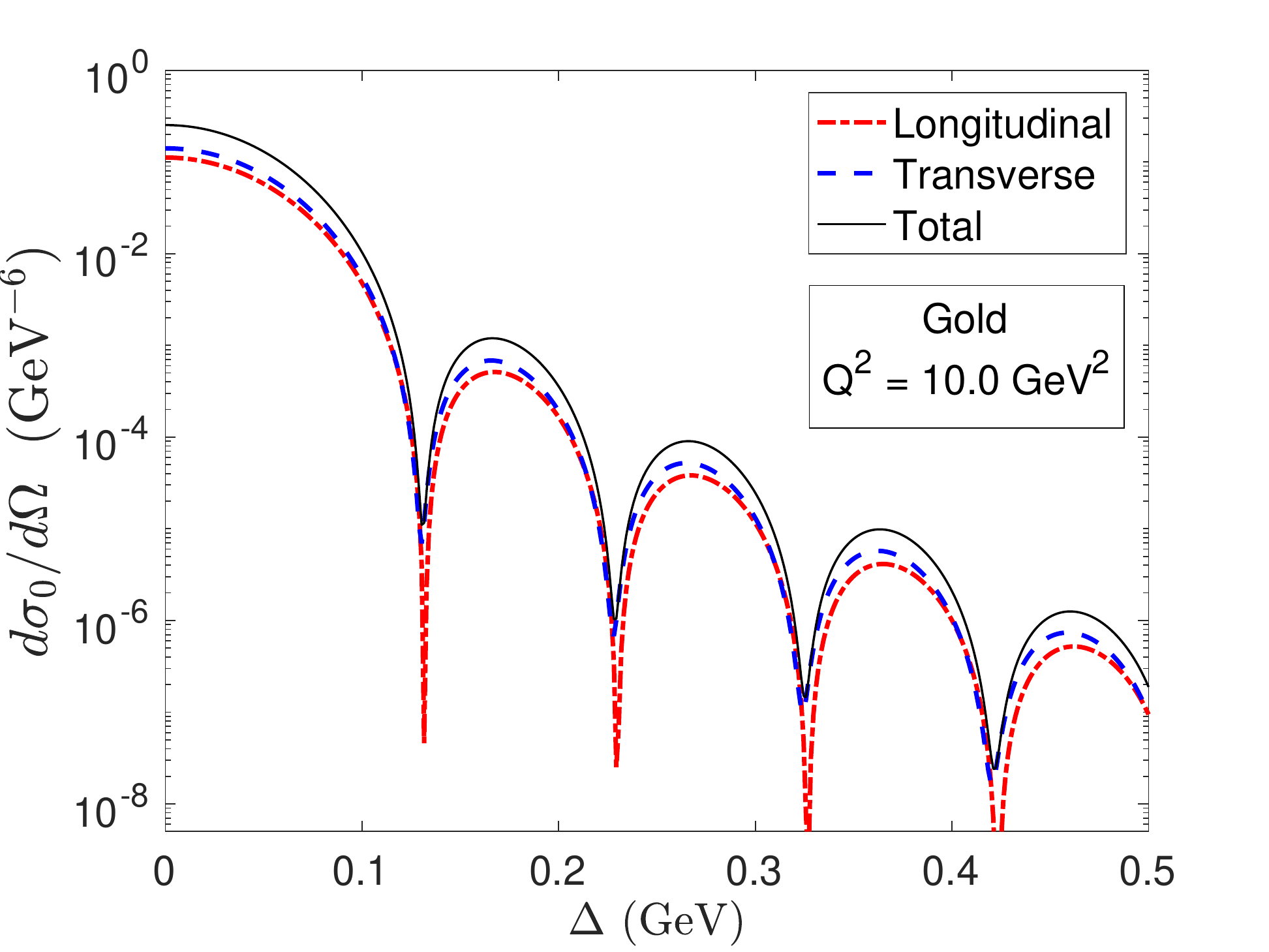}
        \caption{Differential cross section $d \sigma_0 / d \Omega$ as a function of $\Delta$ for a gold target. Here $P = 1.0$ GeV and $Q^2= 10.0$ GeV$^2$.}  \label{sigma_0_Deltadep_Au_Q10}
    \end{minipage}
\end{figure*}

\begin{figure*}[!htb]
    \centering
    \begin{minipage}[t]{0.45\textwidth}
        \centering
        \includegraphics[width=8cm]{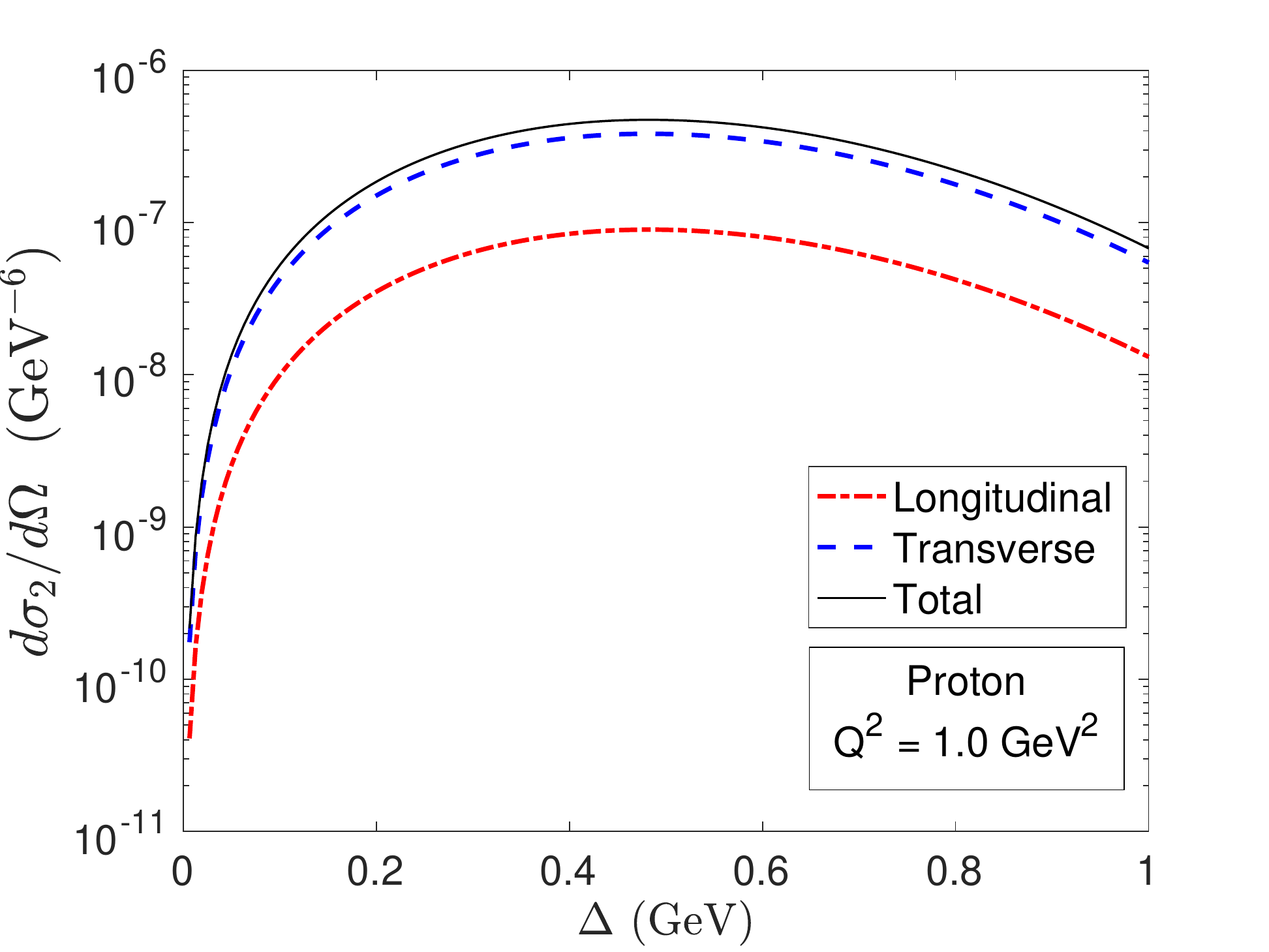}
        \caption{Elliptic anisotropy $d \sigma_2 / d\Omega$ as a function of $\Delta$ for a proton target. Here $P= 1.0$ GeV and $Q^2= 1.0$ GeV$^2$.} \label{sigma_2_Deltadep_proton_Q1}
        \includegraphics[width=8cm]{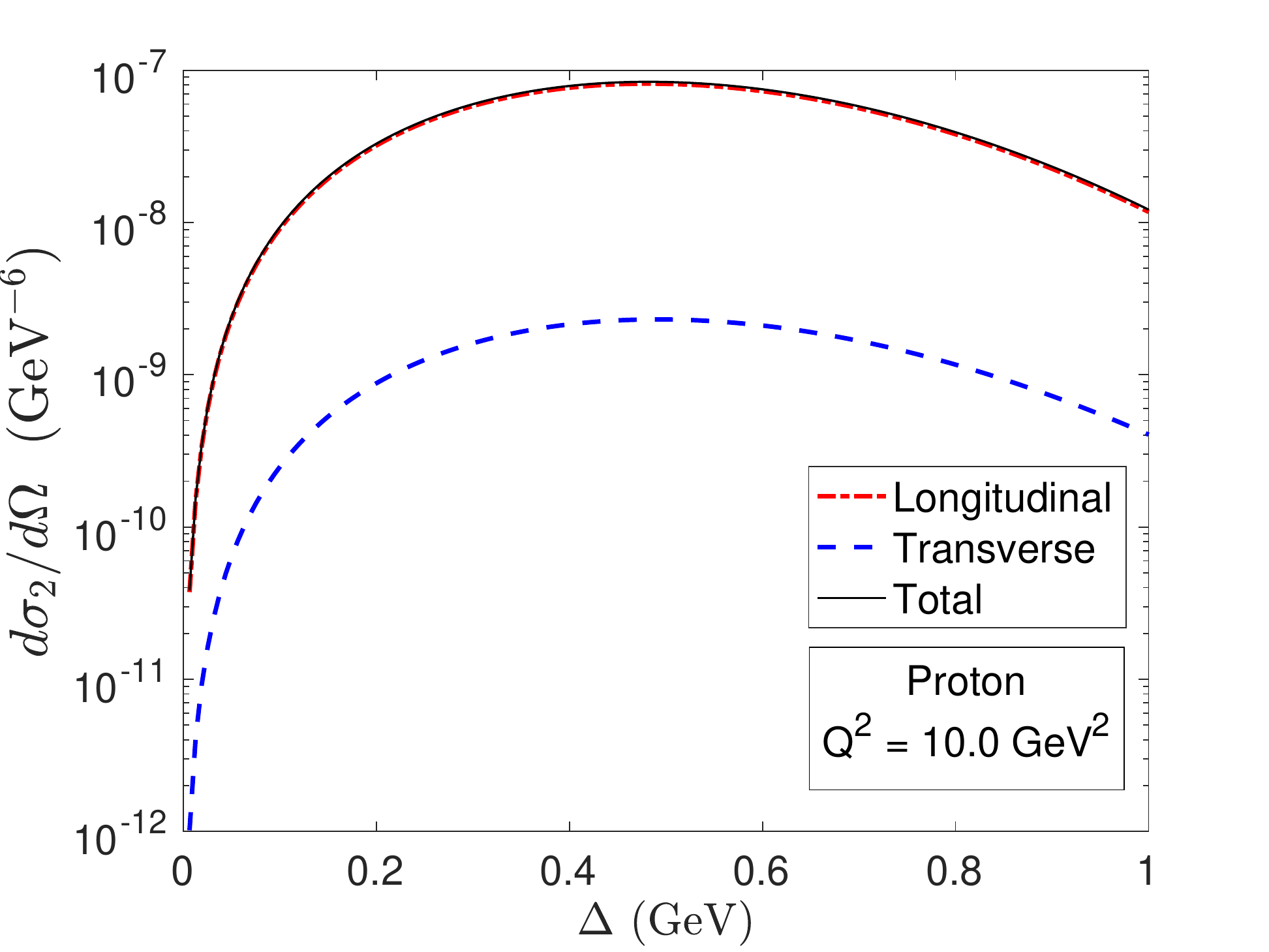}
        \caption{Elliptic anisotropy $d \sigma_2 / d\Omega$ as a function of $\Delta$ for a proton target. Here $P= 1.0$ GeV and $Q^2= 10.0$ GeV$^2$.}  \label{sigma_2_Deltadep_proton_Q10}
    \end{minipage}
    \hspace{0.5cm}
    \begin{minipage}[t]{0.45\textwidth}
        \centering
        \includegraphics[width=8cm]{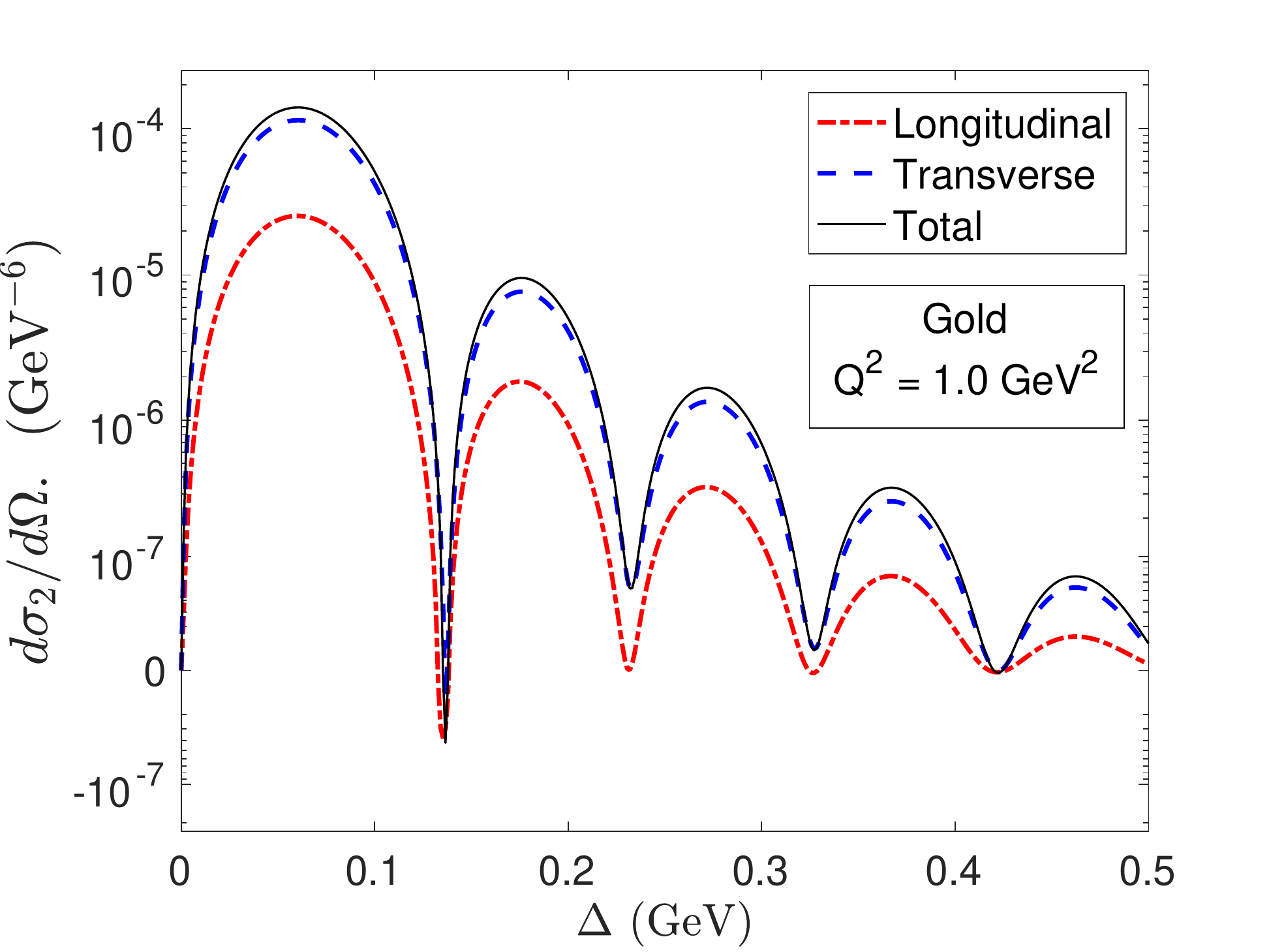}
        \caption{Elliptic anisotropy $d \sigma_2 / d\Omega$ as a function of $\Delta$ for a gold target. Here $P= 1.0$ GeV and $Q^2= 1.0$ GeV$^2$.}  \label{sigma_2_Deltadep_Au_Q1}
        \includegraphics[width=8cm]{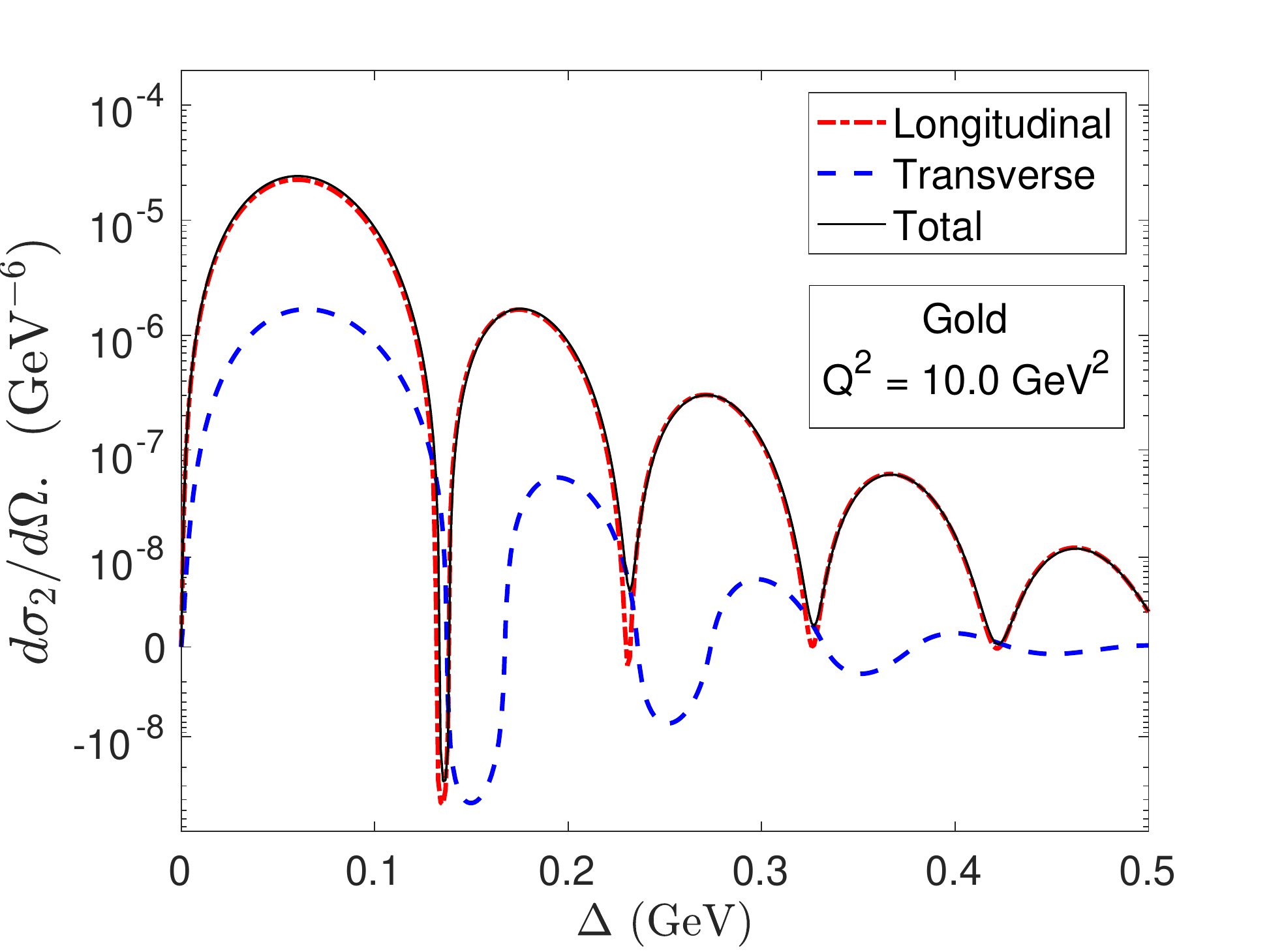}
        \caption{Elliptic anisotropy $d \sigma_2 / d\Omega$ as a function of $\Delta$ for a gold target. Here $P= 1.0$ GeV and $Q^2= 10.0$ GeV$^2$.} \label{sigma_2_Deltadep_Au_Q10}
    \end{minipage}
\end{figure*}

\subsubsection{Approaching the saturated regime $Q_s \sim \varepsilon_f$}
\label{sec:IPMVanalyticSaturation}

Similarly to the case of the GBW-model, we expect the $P$-dependence and $\Delta$-dependence of the differential cross section and elliptic anisotropy not to factorize, and their growth with $Q_s$ to slow down, as we move away from the limit discussed above. Compared to that limit, the $P$-dependent longitudinal differential cross section will develop a dip at larger values of $P$, and the elliptic anisotropy will change sign at larger values of $P$ as well. 

The conditions for the elliptic anisotropy for the transversely polarized photon in Eqs.\,\eqref{Cond1} and \eqref{Cond2} will be modified as $Q_s$ increases. The numerics will show that the change in sign in this case will happen for larger virtuality $Q^2$. The $\Delta$-dependence of the differential cross section and elliptic anisotropy will encode both the geometry of the target color charge profile and the saturation scale, both interlaced by the exponent in the dipole amplitude. These effects, due to the emergence of saturation, will be studied numerically in the next section.  

\section{Numerical results}\label{sec:numrel}

In this section we numerically evaluate the semi-analytic formulas for the differential cross section $d\sigma_0 / d\Omega$ and the elliptic anisotropy $d\sigma_2 / d\Omega$ in Eqs.\,\eqref{CrossLFourier} and \eqref{CrossTFourier} using the dipole amplitude of the impact parameter dependent MV model of Eqs.\,\eqref{IPMVnoangular} and \eqref{IPMVangular} together with the projection formulas in  Eq.\,\eqref{Dprojections}. In the first two subsections we study the $P$ and $\Delta$ dependence. We perform our analysis for both protons and gold nuclei, and two different photon virtualities, $Q^2 = 1\,{\rm GeV^2}$  and $10\, {\rm GeV^2}$. We further employ $m_f=1.28$ GeV for the mass of the charm quark. We choose  $m=$  0.4 GeV as our infrared regulator. For simplicity, our analysis is performed at fixed $z_0=z_1=$ 0.5, which dominates the bulk of the cross section and could be fixed in experiments as well.

For our study of the proton we use $Q^2_{sp}= $ 0.5 GeV$^2$ and a Gaussian target profile
\begin{align} \label{protonprofile}
T_p(b) = e^{-b^2/(2R_p^2)}\,,
\end{align}
with $R_p= 0.4\,{\rm  fm}$, which is the gluonic radius of the proton. The normalization is chosen such that $T_p(b)=1$ at $b=0$. In this case the value of $Q_s$ quoted is that in the center of the proton.
 
For our analysis of the nucleus the thickness function $T_A$ in the transverse plane is obtained by the integration of a Woods-Saxon distribution along the longitudinal direction
\begin{align} \label{nuclearprofile}
T_{A}(b) = N_A \int dz \rho_A(\sqrt{b^2+z^2})\,,
\end{align}
where the Woods-Saxon distribution is given by
\begin{align} \label{nucleardensity}
\rho_A(r) = \frac{1}{1+e^{(r-R_A)/a_A}}\,,
\end{align}
and $N_A$ is chosen such that $T_A(0)=1$.
For a gold nucleus ($A=197$), we choose $R_A = 6.37$ fm and $a_A =  0.535$ fm. The saturation scale is $Q^2_{sA}=1.09 \, {\rm GeV}^2$. Once again, the normalization is such that the $Q_s$ quoted is that in the center of the nucleus. The relation between proton saturation $Q^2_{sp}$ and nuclear saturation scale $Q^2_{sA}$ is discussed in Appendix \ref{app:QsNuclear}.

We will discuss how to use a varying mass number and Bjorken-$x$ to uncover effects of saturation in the differential dijet cross section in Section \ref{sec:saturatedregime}. 

\subsection{$P$ dependence}
\label{sec:Pdep}

In this section we study the $P$ dependence of the differential cross section and elliptic anisotropy at fixed momentum transfer $\Delta = 0.1$ GeV. In all plots we display the longitudinal, transverse, and total cross sections separately.

\subsubsection{Differential cross section $d \sigma_0 /d \Omega$}

We first study the differential cross sections, shown in Figs.\,\ref{sigma_0_pdep_proton_Q1}, \ref{sigma_0_pdep_proton_Q10}, \ref{sigma_0_pdep_Au_Q1}, and \ref{sigma_0_pdep_Au_Q10}. The most salient feature is the dip in the longitudinal cross section in all four cases. The location of the dip shifts to larger momentum with increasing photon virtuality $Q^2$. The location of the dip is at larger momentum $P$ for the gold target, compared to the proton, because the saturation scale $Q_s$ is larger for the former, and in general both $Q_s$ and $Q^2$ affect the $P$-dependence of the cross section, as discussed in the previous section.

In all four cases the transverse components dominate for most values of $P$. However, we see that the difference between the longitudinal and transverse cross sections decreases with increasing photon virtuality. At small $P$ the longitudinal component dominates for $Q^2=10\,{\rm GeV}^2$ 

The cross sections decrease with photon virtuality $Q^2$ as this restricts the size of the dipoles contributing to the scattering. Furthermore, they are more than two orders of magnitude larger for gold nuclei compared to protons, because of the larger size of the target and the increased saturation scale ($Q_s^2\sim A^{1/3}$).

Observe that despite differences in the specific details, i.e., the precise location of the dip, ordering of transverse to longitudinal components, etc., the structure of the results is similar in all four cases, owing to the fact that the $P$-dependence is most sensitive to the form of the light-cone wave functions (projectile) and less sensitive to the target under consideration. 

\subsubsection{Elliptic anisotropy $d \sigma_2 /d \Omega$}

The elliptic anisotropy $d\sigma_2/d\Omega$ is shown in Figs.\,\ref{sigma_2_pdep_proton_Q1}, \ref{sigma_2_pdep_proton_Q10}, \ref{sigma_2_pdep_Au_Q1}, and \ref{sigma_2_pdep_Au_Q10}, using a symmetrical logarithmic scale to display both negative and positive values.

The magnitude of the elliptic anisotropy follows a similar pattern to that of the differential cross section, decreasing with increasing $Q^2$, and increasing with larger target size and saturation scale $Q_s$. In all four cases one observes a maximum at $P\approx 1.0$ GeV, with only a weak dependence of this location on virtuality or saturation scale (which varies with the choice of target). As expected the elliptic anisotropy vanishes at large values of $P$ and at $P=0$ GeV.

For the longitudinally polarized photon, the elliptic anisotropy changes sign in all four cases. As anticipated from our analytic investigation, this happens at values of $P$ that coincide with the location of the dip in $d \sigma_0/ d \Omega$, which shifts to larger values of $P$ as the virtuality $Q^2$ and saturation scales $Q_s$ increase.

For the transversely polarized case, at low virtuality $Q^2$ the elliptic anisotropy remains positive (Figs.\,\ref{sigma_2_pdep_proton_Q1} and \ref{sigma_2_pdep_Au_Q1}), while at large virtuality $Q^2$, the elliptic anisotropy changes sign (Fig.\,\ref{sigma_2_pdep_proton_Q10}), as discussed in the last paragraph of Section \ref{sec:IPMVcross}. Since $Q_s$ is larger for the gold nucleus, the used $Q^2$ is not large enough to cause a change of sign in the result shown in Fig.\,\ref{sigma_2_pdep_Au_Q10}. However, as one increases the virtuality further the anticipated sign change will appear: We have checked that $d\sigma_{T,2}/d\Omega$ changes sign as a function of $P$ for $Q^2 \gtrsim 12 \,{\rm GeV}^2$.

The transversely polarized contribution to the elliptic anisotropy dominates at low $Q^2$, but at high $Q^2$ we observe that the longitudinal piece starts to dominate for values of $P$ below 2 GeV.

The order of magnitude for the relative total elliptic anisotropy is about $0.1\%$ for the chosen value of $\Delta=0.1\,{\rm GeV}$, in agreement with the results in \cite{Mantysaari:2019csc}.

\subsection{$\Delta$ dependence}
\label{sec:Deltadep}

In this section we study the $\Delta$ dependence of the differential cross section  $d \sigma_0 /d \Omega$ and elliptic anisotropy  $d \sigma_2 /d \Omega$ at fixed $P = 1.0$ GeV. In all plots we display the longitudinal, transverse, and total cross sections separately.

\subsubsection{Differential cross section $d \sigma_0 /d \Omega$}
\label{sec:delta_elliptic}
We present the $\Delta$-dependence of the cross section for diffractive dijet production off a proton in Figs.\,\ref{sigma_0_Deltadep_proton_Q1} and \ref{sigma_0_Deltadep_proton_Q10} for $Q^2=1$ and 10 GeV$^2$, respectively. In both cases one observes diffractive dips at values comparable to the inverse size of the proton $1/R_p$. This is not purely a geometric effect, but relies on the presence of saturation. For example, the results cannot be anticipated from the non-saturated regime (quadratic expansion of the dipole amplitude, e.g. Eqs.\,\eqref{ToyIPMVsigma0L} and \eqref{ToyIPMVsigma0T}), where one would obtain a smooth $\Delta$-dependence from the used Gaussian profile in coordinate space. In fact, this behavior cannot be obtained in any finite order Taylor expansion of the dipole amplitude. It is a consequence of the (resummed) multiple scattering of the dipole, or unitarity of the dipole amplitude \cite{Armesto:2014sma,Altinoluk:2015dpi}.
Consequently, the location of the dip depends on $Q_s$, which we will demonstrate explicitly in the next subsection. As before, one further observes in Figs.\,\ref{sigma_0_Deltadep_proton_Q1} and \ref{sigma_0_Deltadep_proton_Q10} that with increasing $Q^2$ the difference between cross sections for transverse and longitudinal polarizations decreases.

Results for a gold nucleus target are presented in Figs.\,\ref{sigma_0_Deltadep_Au_Q1}, and \ref{sigma_0_Deltadep_Au_Q10}. The various diffractive dips occur at multiples of $\sim 3/R_A \sim 0.1$ GeV (the factor of 3 is closely related to the zeroes of the Bessel function $J_0$), where $R_A$ is the size of the nucleus. The locations of the diffractive dips also depend on $a_A$ (the skin depth of the nucleus). Unlike for the (Gaussian) proton, these dips are present due to the geometry of the nucleus. Nonetheless, their locations are also modified by the unitarization of the dipole amplitude. We have checked that in the absence of saturation (expansion to quadratic order), the dips shift to larger values of $\Delta$. At large $Q^2$ the difference between the longitudinal and transverse component of the differential cross section is reduced.

\begin{figure*}[!htb]
    \centering
    \begin{minipage}[t]{0.45\textwidth}
        \centering
        \includegraphics[width=8.5cm]{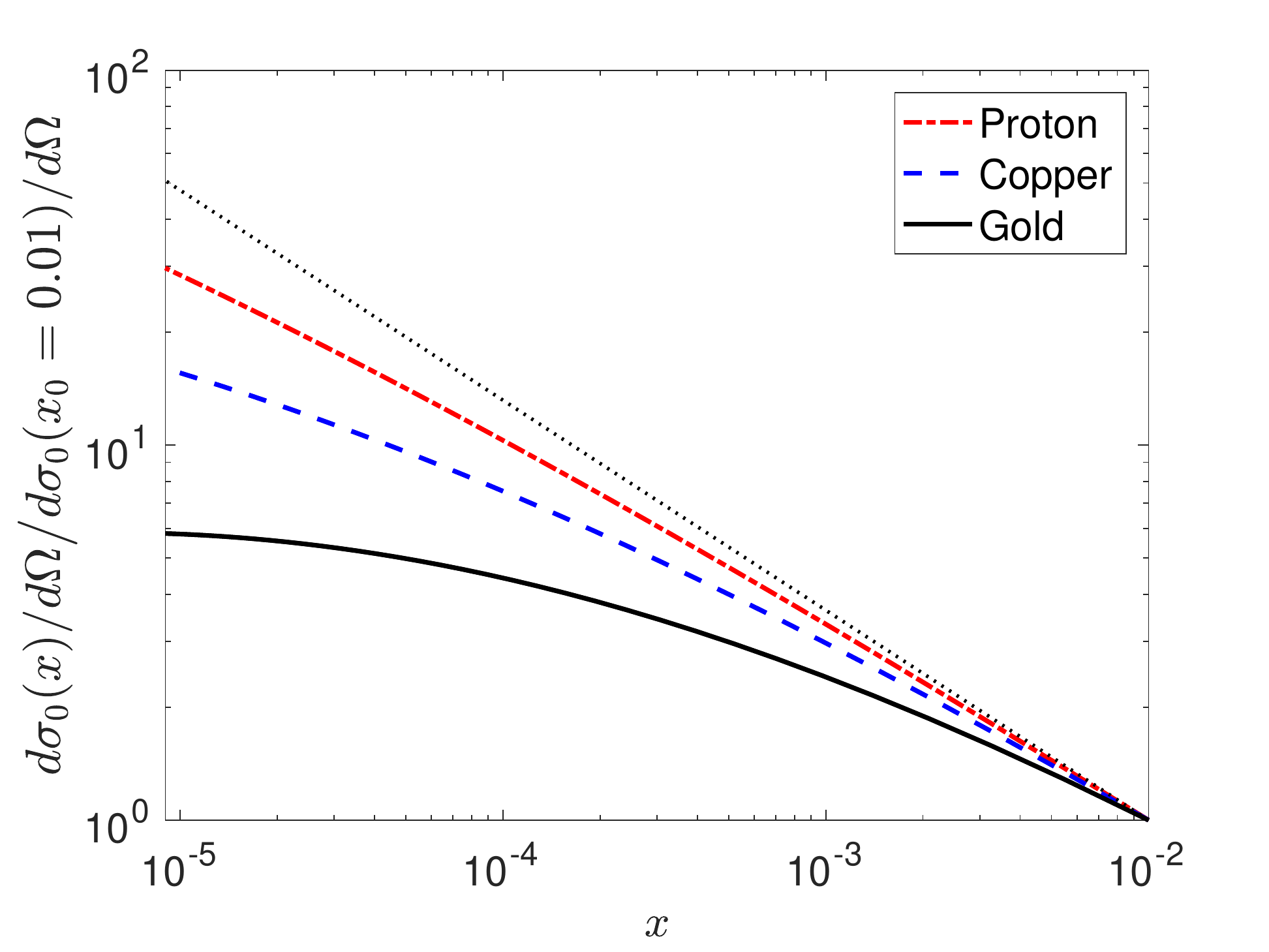}
        \caption{Evolution of the differential cross section (here $P$ = 1.0 GeV, and $\Delta$ = 0.1 GeV) with $x$ for proton, copper and gold.} \label{P_plot_x_and_A}
    \end{minipage}
    \hspace{0.5cm}
    \begin{minipage}[t]{0.45\textwidth}
        \centering
        \includegraphics[width=8.5cm]{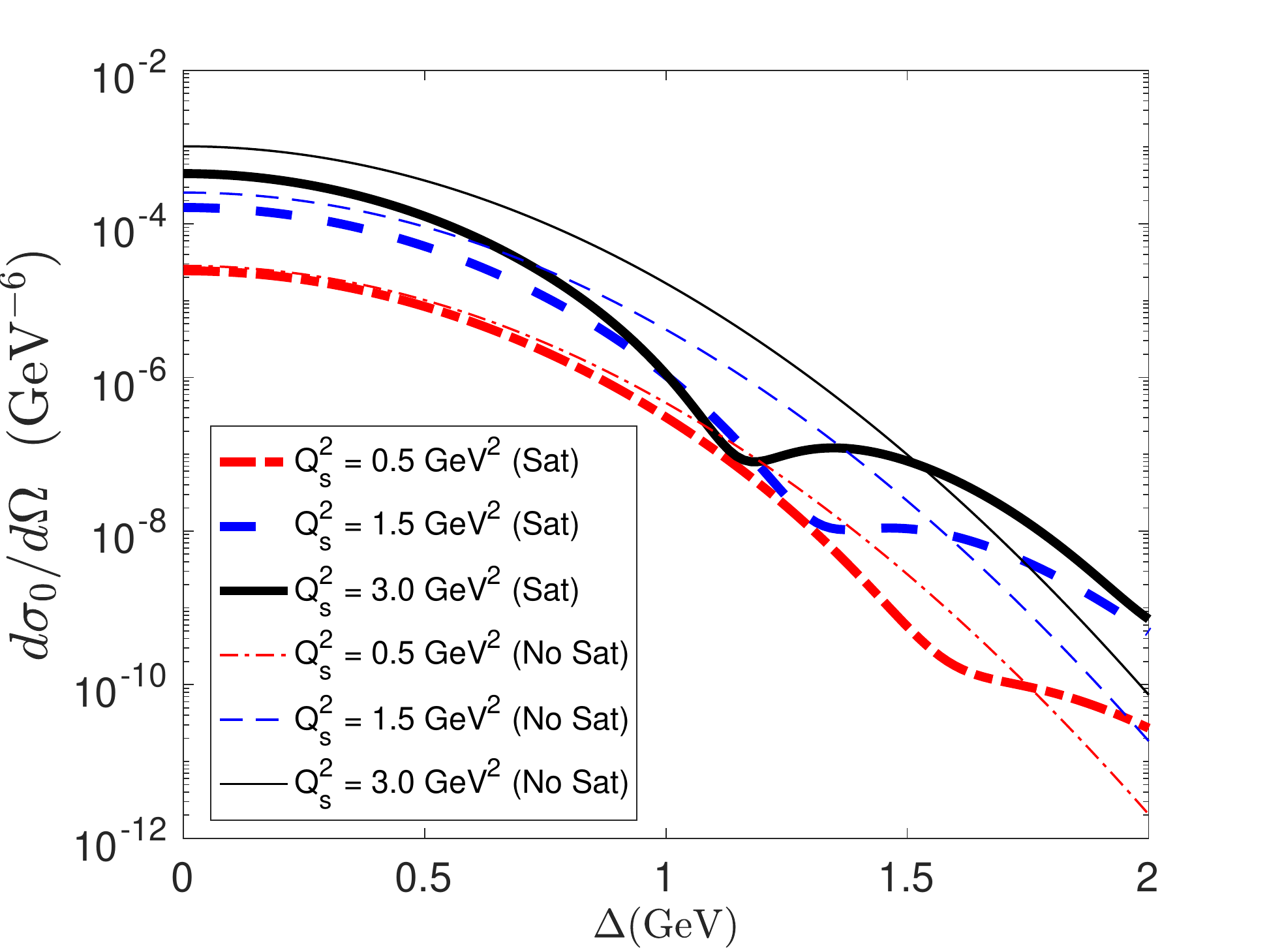}
        \caption{Target: proton. Differential cross section as a function of $\Delta$, for different values of $Q^2_s$. Here $P= 1.0$ GeV, and $Q^2$ = 1.0 GeV$^2$.}  \label{Delta_plot_Qs}
    \end{minipage}
\end{figure*}

\subsubsection{Elliptic anisotropy $d \sigma_2 /d \Omega$}

The dependence of the elliptic anisotropy on the momentum transfer $\Delta$ is shown for a proton target in Figs.\,\ref{sigma_2_Deltadep_proton_Q1} and \ref{sigma_2_Deltadep_proton_Q10} for $Q^2=1$ and $10\, {\rm GeV}^2$, respectively. Unlike the differential cross section, here we only show results for $\Delta \leq 1.0$ GeV. This is because the validity of our approximation breaks down for the elliptic anisotropy at large $\Delta$ (see the discussion at the end of Appendix \ref{app:ipmv}). The anisotropy  increases rapidly with $\Delta$ and reaches a maximum at $\Delta \approx 0.5 $ GeV, which is approximately the inverse gluonic size of the proton. At $Q^2=10$ GeV$^2$, the longitudinal component dominates the elliptic anisotropy, while at the smaller $Q^2=1\,{\rm GeV}^2$ the transverse component dominates. This is in agreement with Figs.\,\ref{sigma_2_pdep_proton_Q1} and \ref{sigma_2_pdep_proton_Q10}.

We conclude our discussion with the gold nucleus, shown in Figs.\,\ref{sigma_2_Deltadep_Au_Q1} and \ref{sigma_2_Deltadep_Au_Q10}. The behavior resembles that of the differential cross section, except that a global maximum appears at $\Delta \sim 0.05$ GeV, which is close to the inverse size of the nucleus. An interesting feature is the presence of regions where the elliptic anisotropy is negative. This is an effect of the unitary (exponentiated scattering) of the dipole amplitude, as such regions will not be present if the the dipole was expanded to quadratic order. This can be seen directly in Eqs.\,\eqref{ToyIPMVsigma2L} and \eqref{ToyIPMVsigma2T}, which show that the elliptic anisotropy is proportional to $\Delta^2 |\tilde{T}(\Delta)|^2$, meaning that no sign change will occur with varying $\Delta$. 

Our results on the $P$- and $\Delta$-dependence of the elliptic anisotropy presented above provide important guidance for future experiments. Note that the elliptic anisotropy varies strongly as a function of both $P$ and $\Delta$. Consequently, the choice of the kinematics of the dijet will affect how well the anisotropy can be measured experimentally. For example, for a proton target, choosing $P\approx 1\,{\rm GeV}$ and $\Delta\approx 0.5\,{\rm GeV}$ is predicted to maximize the elliptic anisotropy, while for a gold target a smaller $\Delta\approx 0.05\,{\rm GeV}$ would be preferred.
 
\subsection{Observable effects of approaching the saturated regime}
\label{sec:saturatedregime}

In this section we study effects of saturation that are potentially observable in measurements of diffractive dijet production at a future electron ion collider. We focus our attention on the differential cross section.

In Fig.\,\ref{P_plot_x_and_A} we show the $x$-dependence of the differential cross section at fixed values of $P$ = 1.0 GeV and $\Delta = $ 0.1 GeV for different targets: proton, copper ($R_A=4.163\,{\rm fm}$, $a_A=0.606$), and gold. We normalized the differential cross sections by their values at $x_0 = 0.01$. In the semi-analytic model used here, the evolution in $x$ solely affects the value of $Q_s$. We thus simply vary $Q_s$ and relate it to $x$ using the parametric relation
\begin{align} \label{Qs_scaling_x}
    \frac{Q^2_s}{Q^2_{s0}} = \left( \frac{x_0}{x} \right)^{0.28}\,.
\end{align}
For the initial saturation scales $Q^2_{s0}$ of the proton, copper and gold nucleus we chose: 0.3, 0.43, 0.65 GeV$^2$, respectively. The relation between the proton saturation scale $Q^2_{sp}$ and the nuclear saturation scale $Q^2_{sA}$ is given in Appendix \ref{app:QsNuclear}.  

The reference dotted line shows the expected evolution in the absence of saturation. In that case, the differential cross section grows with $Q_s^4$ (see Eqs.\eqref{ToyIPMVsigma0L} and \eqref{ToyIPMVsigma0T}). The results show the slowdown of the growth of the differential cross section in response to saturation effects. These set in earlier for the denser targets because of their larger saturation scale $Q_s^2 \sim A^{1/3}$ for any given $x$. C.f. the discussion in Sec. \ref{sec:IPMVanalyticSaturation}.

In Fig. \ref{Delta_plot_Qs} we show the differential cross section as a function of $\Delta$ for a proton target at different values of the saturation scale $Q_s$. The narrow lines denote a "non-saturation" model where the dipole amplitude is not exponentiated (expanded to quadratic order). The figure shows that the differential cross sections are smaller for the case including saturation (thick lines). This effect is more pronounced for larger saturation scales as expected. 

The more prominent feature is the dependence of the location of the diffractive dip on the value of $Q^2_s$. As the saturation scale increases, the dip shifts to lower values of $\Delta$. This effect has been observed in \cite{Altinoluk:2015dpi} and explained by the fact that the effective spatial shape of the proton is non-Gaussian. This happens because the center of the proton is approaching the black disk limit (mathematically, this happens because the Gaussian thickness functions appears in the exponential of the dipole amplitude). The dip is absent in the non-saturated case, as the differential cross section here is proportional to the square of the Fourier transform of the Gaussian profile, which does not have a dip structure (See Eqs. \eqref{ToyIPMVsigma0L} and \eqref{ToyIPMVsigma0T}). 

If the $\Delta$-dependent cross section could be measured at a future electron ion collider for different values of $x$, and a similar systematic change of the dip position observed, it would be an interesting indication that we are approaching the saturated regime. Note that of course there are some caveats, since the detailed shape of the proton is not known and here we do not consider the growth of the proton with decreasing $x$ \cite{Gribov:1973jg,Kovner:2001bh,Schlichting:2014ipa,Mantysaari:2018zdd}, which will likely affect the detailed quantitative result. 

\section{Conclusions}\label{sec:conclusions}

We have studied analytically and semi-analytically the properties of coherent diffractive dijet production in electron-proton and electron-nucleus collisions, using two different saturation models including impact parameter (and angular) dependence: the Golec-Biernat Wusthoff model and an Impact Parameter dependent McLerran Venugopalan model.

We derived general relations connecting angular correlations of the dipole orientation and impact parameter vector in coordinate space with angular correlations between mean dijet transverse momentum and hadron recoil momentum (Eq.\,\eqref{ModeRelation}). We showed that the $n-$th Fourier harmonic of the amplitude for diffractive dijet production (in momentum space) depends only on the $n-$th harmonic of the dipole amplitude in coordinate space.

In the limit of large photon virtuality $Q^2$ and/or quark mass, the differential cross sections and elliptic anisotropies of the GBW model can be expressed completely analytically. In this limit, the $P$- and $\Delta$-dependencies factorize, providing distinct information on the projectile and target. The $P$-dependence showed interesting analytic structures such as a dip in the differential cross section (for longitudinal photon polarization), and changes in sign of the elliptic anisotropy (for both longitudinal and transverse polarizations), and provided insight into how these features depend on $Q^2$, the quark mass, and the longitudinal momentum fractions of the quark and anti-quark.
The $\Delta$-dependence directly probes features of the target, being sensitive to the Fourier transform of the transverse density distribution.

In the case of the more realistic IP-MV model, where the anisotropy is explicitly driven by the gradients of the target geometry, we found approximate analytic expressions for the differential cross sections and elliptic anisotropies. In particular, we observed that the $P$-dependence was modified from the GBW model because of the presence of the logarithm in the dipole amplitude. Both the locations of dips in the longitudinal cross section and sign-change in the elliptic anisotropies shifted to larger values of $P$.

Approaching the saturation limit, we discussed the expected  modification to the features mentioned above; in particular their dependence on the saturation scale $Q_s$. A more detailed analysis of the effects of approaching saturation could only be performed by numerical evaluation of our semi-analytic expressions.

The numerical results confirmed several expectations. We observed an increase of the value of $P$ at the dip position in the longitudinal differential cross section with increasing photon virtuality $Q^2$ and saturation scale $Q_s$ (increasing mass number $A$). A similar behavior was found for the change in sign in the elliptic anisotropy (for longitudinal photons). At low $Q^2$ the transversely polarized differential cross section and elliptic anisotropy dominate over their longitudinally polarized counterparts. The difference between them decreases with increasing $Q^2$.

As a function of momentum transfer $\Delta$, we observed the anticipated diffractive dips in the differential cross section for gold. We also observed a dip in the $\Delta$-dependent cross section for a proton target, despite the Gaussian shape of the assumed proton density profile. In the latter case, dips appear because of the unitarization of the dipole amplitude, signaling the effects of saturation, which leads to an effectively non-Gaussian shape of the proton. The $\Delta$-dependence of the elliptic anisotropy also showed some effects of saturation such as the change in sign for the case of gold.

To gain more insight into the effects of saturation, we studied the $x$-evolution of the differential cross section at fixed values of $P$ and $\Delta$. We observed the expected slow-down in the growth of the differential cross section with decreasing $x$, with the effect setting in at larger values of $x$ for the larger targets because of their larger saturation scales at a given $x$. 

Finally, we studied the $\Delta$-dependence of the diffractive dijet cross section for a proton target and different values of the saturation scale (representing varying $x$ values), and observed a decrease of the value of $\Delta$ at the diffractive dip position with increasing  $Q_s$. We compared to the "non-saturation" model, where the dipole was expanded to quadratic order, and did not observe any dips as expected. We argued that if such diffractive dips and their dependence on $x$ could be measured experimentally in diffractive dijet events in e+p collisions, this could provide a strong indication of the presence of saturation effects.

The semi-analytic approach presented in this paper does not include some potentially important physical effects of the small $x$ evolution, such as the growth of the color charge profile with decreasing $x$, and the corresponding modification of the color charge density gradients. In this work the $x$-evolution was only incorporated via the parametrization of the saturation scale $Q_s$ in Eq.\,\eqref{Qs_scaling_x}. The effect was included in numerical studies using g Jalilian Marian-Iancu-McLerran-Weigert-Leonidov-Kovner evolution \cite{JalilianMarian:1996xn,JalilianMarian:1997jx,JalilianMarian:1997gr,Iancu:2000hn,Iancu:2001ad,Iancu:2001md,Ferreiro:2001qy}, and has been analyzed in \cite{Mantysaari:2019csc}. The evolution of the dipole could also be studied using the BK evolution \cite{Balitsky:1995ub,Kovchegov:1999yj} with impact parameter dependence as studied in \cite{Berger:2011ew}. Also, a more detailed analysis will require the incorporation of the dependence of $x$ on the dijet momenta \cite{Mantysaari:2019csc}.

Nevertheless, what makes our approach a very powerful tool for understanding what physical features of projectile and target are important for the process of diffractive dijet production in e+p and e+A collisions, is that we were able to find fully analytic expressions for cross sections and elliptic anisotropies in certain limits. 
Even the semi-analytic expressions only involve simple integrals that are easily evaluated numerically, which is especially helpful for examining the regimes of large $\bm{P}$ and/or $\bm{\Delta}$.
This allows us to efficiently constrain the most interesting setup and kinematic regions in future experiments. In particular, we provided predictions for values of $P$ and $\Delta$ that maximize the magnitude of the elliptic anisotropies for different targets to assist future experiments in observing these interesting correlations.

Finally, we point out that the presented techniques could be extended to study other processes such as inclusive dijet production. In this case, one needs to work out an expression for the quadrupole, incorporating the effects of the geometry of the target. 
Similarly, one could attempt to extend this analysis to inclusive diffractive dijet and incoherent diffractive dijet production. In the latter case we expect to gain sensitivity to the local structures of the target and possibly angle dependent fluctuations.

\section*{Acknowledgements}
We thank Renaud Boussarie, Edmond Iancu, Heikki M\"antysaari, Niklas Mueller, Alba Soto-Ontoso, Derek Teaney and Raju Venugopalan for useful discussions. 
FS and BPS are supported by the U.S. Department of Energy, Office of Science, under contract No. DE-SC0012704.

\appendix

\section{Angular correlations: from coordinate space to momentum space} \label{app:moderelation}

In this appendix we prove the relation in Eq.\,\eqref{ModeRelation}. We first lay out the conventions for the Fourier transforms and mode expansion. The Fourier transform and inverse Fourier transform are normalized as follows
 \begin{align}
\tilde{F}(\bm{P},\bm{\Delta}) = &\int d^2\bm{r} d^2\bm{b} \  e^{-i \bm{P} \cdot \bm{r}} e^{-i \bm{\Delta} \cdot \bm{b}} F(\bm{r},\bm{b}) \label{FourierT} \,, \\
F(\bm{r},\bm{b}) =& \int \frac{d^2\bm{P}}{(2\pi)^2} \frac{d^2\bm{\Delta}}{(2 \pi)^2} \  e^{i \bm{P} \cdot \bm{r}} e^{i \bm{\Delta} \cdot \bm{b}} \tilde{F}(\bm{P},\bm{\Delta}) \,. \label{FourierTInv}
\end{align}
For rotationally symmetric functions, the Fourier mode decomposition is given by
\begin{align}
F(r,b,\theta_{rb}) &= F_0(r,b) + 2 \sum_{k=1}^{\infty} F_k(r,b) \cos \left( k  \theta_{rb} \right) \,, \label{FTECoord}  \\
\tilde{F}(P,\Delta,\theta_{P\Delta}) &= \tilde{F}_0(P,\Delta) + 2 \sum_{k=1}^{\infty} \tilde{F}_k(P,\Delta) \cos \left( k  \theta_{P\Delta} \right) \,, \label{FTEMome}
\end{align}
where the $F_k$ and $\tilde{F}_k$ can be computed by projection
\begin{align}
F_k(r,b) &= \frac{1}{2\pi} \int_0^{2\pi} d\theta F(r,b,\theta) \cos( k\theta) \,, \nonumber \\
\tilde{F}_k(P,\Delta) &= \frac{1}{2\pi} \int_0^{2\pi} d\theta \tilde{F}(P,\Delta,\theta) \cos( k\theta) \,. \label{FTProjection} 
\end{align}
To prove Eq.\,\eqref{ModeRelation}, we use the following identity
\begin{align}
e^{-iA\cos\phi} = \sum_{n=-\infty}^\infty (-i)^n J_n(A)e^{-i n\phi} \,.
\end{align}
Then the Fourier transform Eq.\,\eqref{FourierT} can be expressed as a Bessel expansion. Changing the variables to $\theta_{rb}$ and $\Theta = \frac{1}{2} \left(\theta_r + \theta_b \right)$, we arrive at
\begin{align}
\tilde{F}(\bm{P},\bm{\Delta}) =&\int r dr  b db  \sum_{n,m=-\infty}^\infty (-i)^{n+m} \notag\\ &\times J_n(Pr) J_m(\Delta b) e^{-i\left( n\theta_P +m \theta_\Delta  \right) }  \notag \\ & \times \int d \Theta d\theta_{rb} e^{i\left(  (n+m) \Theta + \frac{1}{2}(n-m) \theta_{rb} \right) }  F(r,b,\theta_{rb})
\end{align}

The $\Theta$ integral is trivial and is proportional to a Kronecker-Delta, $2 \pi \delta_{n,-m}$, which we use to contract the summation in $m$, yielding
\begin{align}
\tilde{F}(\bm{P},\bm{\Delta}) =& 2\pi  \int rdr  b db  \sum_{n=-\infty}^\infty (-1)^n J_n(Pr) J_n(\Delta b) \notag \\ & \times e^{-in \left(  \theta_P -\theta_\Delta  \right) }  \notag\\ & \times \int  d\theta_{rb} e^{in  \theta_{rb} }  F(r,b,\theta_{rb}) \,.
\end{align}
where we used $J_{-n}(z) = (-1)^n J_n(z)$.

To perform the angular integral, we plug in the Fourier mode expansion Eq.\,\eqref{FTECoord} and find
\begin{align}
\tilde{F}(\bm{P},\bm{\Delta})  = (2\pi)^2  \int rdr  b db  \sum_{n=-\infty}^\infty (-1)^n J_n(Pr) J_n(\Delta b)  \nonumber \\ \left(F_0(r,b)  \delta_{n0}  + 2\sum_{k=1}^\infty \delta_{nk} F_k(r,b) \cos (k \theta_{P\Delta}) \right) \,.
\end{align}
Contracting the summation in $n$ and comparing with the expansion in Eq.\,\eqref{FTEMome} we find
\begin{align}
\frac{\tilde{F}_k(P,\Delta)}{(2\pi)^2} = (-1)^k \int rdr  \ b db \ J_k(Pr) J_k(\Delta b) F_k(r,b) \,.
\end{align}

\section{Useful integral identities}
\label{app:integrals}
Representation of Bessel functions of the first kind
\begin{align}
J_0(z) &= \frac{1}{2\pi}\int_0^{2\pi} d\phi e^{-iz \cos \phi} \,,  \label{BesselJ}\\
J_2(z) &=-\frac{1}{2\pi}\int_0^{2 \pi} d\phi e^{-iz \cos \phi} \cos 2\phi  \,. \label{BesselJ2}
\end{align}
Representation of modified Bessel functions of the first kind
\begin{align}
I_0(z) &= \frac{1}{2 \pi}\int_0^{2\pi} d\phi e^{-z \cos 2\phi} \,, \nonumber \\  I_1(z) &= -\frac{1}{2 \pi} \int_0^{2\pi} d\phi e^{-z \cos 2\phi} \cos 2 \phi\,. \label{BesselI}
\end{align}
The following integral is the backbone for our analytic computations:
\begin{align}
\int rdr J_0(Pr) K_0(\varepsilon_f r) = \frac{1}{P^2 +\varepsilon^2_f} \,. \label{backbone}
\end{align}
By taking derivatives with respect to $P$ or $\varepsilon_f$, and using recurrence relations for derivatives of $J_n(z)$ and $K_n(z)$, one finds
\begin{align}
\int rdr J_0(Pr) r^2 K_0(\varepsilon_f r) &=- \frac{4( P^2 - \varepsilon^2_f )}{(P^2 +\varepsilon^2_f)^3} \,, \nonumber \\
\int rdr J_1(Pr) r^2 K_1(\varepsilon_f r) &= \frac{8P \varepsilon_f }{(P^2 +\varepsilon^2_f)^3} \,, \nonumber \\\
\int rdr J_2(Pr) r^2 K_0(\varepsilon_f r) &= \frac{8 P^2}{(P^2 +\varepsilon^2_f)^3} \,, \nonumber \\
\int rdr \frac{J_3(Pr) - J_1(Pr)}{2} r^2 K_1(\varepsilon_f r) &= \frac{4 P (P^2 - \varepsilon_f^2 )}{\varepsilon_f(P^2 +\varepsilon^2_f)^3} \,. \label{UsefulIntegralr}
\end{align}

It is also useful to have expressions for the Fourier transform of an isotropic function. They follow from the standard definition of the Fourier transform and Eq.\,\eqref{BesselJ}:
\begin{align}
\tilde{T}(\Delta)  &= 2\pi \int b db J_0(\Delta b) T(b) \,,\nonumber \\
T(b) &= \frac{1}{2\pi} \int \Delta \, d \Delta J_0(\Delta b)  \tilde{T}(\Delta) \,. \label{FTisotropic}
\end{align}
One can obtain interesting relations by taking derivatives. For example:
\begin{align}
    \left[\frac{d^2}{db^2} - \frac{1}{b}\frac{d}{db} \right] T(b) = \frac{1}{2\pi} \int \Delta \, d \Delta J_2(\Delta b)  \Delta^2  \tilde{T}(\Delta) \,, \label{IdentityJ2}
\end{align}
where we used
\begin{align}
   \left[\frac{d^2}{db^2} - \frac{1}{b}\frac{d}{db} \right] J_0(\Delta b) &= b \frac{d}{db}\left[ \frac{1}{b} \frac{d}{db} J_0(\Delta b)\right] \nonumber \\
   & =b \frac{d}{db}\left[ -\frac{1}{b} \Delta  J_1(\Delta b) \right]
   \nonumber \\
   &=\Delta^2 J_2(\Delta b) \,. \nonumber 
\end{align}
By inverting Eq.(\ref{IdentityJ2}), one has
\begin{align}
\Delta^2 \tilde{T}(\Delta) = 2\pi \int b db J_2(\Delta b) \left[\frac{d^2}{db^2} - \frac{1}{b}\frac{d}{db} \right] T(b) \,. \label{IdentityJ2inverse}
\end{align}

\section{Details of analytic calculations of differential cross section and elliptic anisotropy}
\label{app:analyticsupplement}
In order to compute the differential cross sections and elliptic anisotropies (Eqs.\,\eqref{CrossLFourier} and \eqref{CrossTFourier}), it is enough to calculate the functions in  Eqs.\,\eqref{ML0}, \eqref{ML2}, \eqref{PNT0}, and \eqref{PNT2}. For the sake of simplicity we ignore the small corrections to the differential cross section, i.e. we only keep the terms $|\tilde{F}_0|^2$ and $|\partial_P \tilde{G}_0|^2$). In this appendix we show the explicit calculations for these expressions in the limit $Q_s \ll \varepsilon_f$, in which the dipole amplitude can be expanded to quadratic order. We start with the GBW model, for which we find exact analytic results, and then proceed to derive approximate expressions for the impact parameter dependent MV model.

\subsection{Golec-Biernat Wusthoff model}
Using Eq.\,\eqref{ML0} with the expanded expression for $D_0$ in Eq.\,\eqref{SmallDipole} we have
\begin{align}
    \tilde{F}_0(P,\Delta) = &\frac{\pi}{2} Q^2_s \int r dr J_0(Pr) r^2 K_0(\varepsilon_f r) \nonumber \\
    &\times  (2\pi)\int b db J_0(\Delta b) T(b) \label{F0GBW} \,.
\end{align}

We solve the integrals in $r$ and $b$ with the help of Eqs.\,\eqref{UsefulIntegralr} and Eqs.\,\eqref{FTisotropic}, respectively, to find
\begin{align}
\tilde{F}_0(P,\Delta) &=  -2\pi Q_s^2 \frac{  \left(P^2 - \varepsilon^2_f \right)}{\left(P^2+\varepsilon^2_f \right)^3} \tilde{T}(\Delta) \,.  \label{F0GBWmodel}
\end{align}
The other expressions can be obtained in a similar fashion to read
\begin{align}
\tilde{F}_2(P,\Delta) &=   2\pi \  \frac{c}{2} \ Q_s^2  \frac{2 P^2}{\left(P^2+\varepsilon^2_f \right)^3} \tilde{T}_2(\Delta) \label{F2GBWmodel} \,, \\
\partial_P \tilde{G}_0(P,\Delta)&= -2\pi Q_s^2  \frac{ 2 P\varepsilon_f  }{\left(P^2+\varepsilon^2_f \right)^3} \tilde{T}(\Delta)  \label{G0GBWmodel} \,, \\
\partial_P \tilde{G}_2(P,\Delta) &= - 2\pi \  \frac{c}{2} \ Q_s^2  \frac{  P(P^2-\varepsilon_f^2)}{\varepsilon_f \left(P^2+\varepsilon^2_f \right)^3} \tilde{T}_2(\Delta) \,, \label{G2GBWmodel}
\end{align} 
where $\tilde{T}_2(\Delta) = 2\pi \int bdb J_2(\Delta b) T(b)$ is the 2nd order Hankel transform of $T(b)$.

\subsection{Impact parameter dependent McLerran Venugopalan model}
We now consider the Impact Parameter dependent McLerran Venugopalan model in the limit $Q_s \ll \varepsilon_f$.
The expressions for $\tilde{F}_2$ and $\partial_P \tilde{G}_2$ can be solved exactly. For example, one has
\begin{align}
    \tilde{F}_2(P,\Delta) =& \frac{2 \pi}{8} Q^2_s \int r dr J_2(Pr) r^2 K_0(\varepsilon_f r) \nonumber \\
    &\times \frac{2\pi}{6m^2}\int b db J_2(\Delta b) \left[\frac{d^2}{db^2} - \frac{1}{b}\frac{d}{db} \right] T(b) \,.
\end{align}
Using Eqs.\,\eqref{UsefulIntegralr} and Eq.\,\eqref{IdentityJ2inverse} to solve the $r$ and $b$ integrals, respectively, we obtain
\begin{align}
    \tilde{F}_2(P,\Delta) & =   2\pi Q_s^2  \frac{ 2P^2}{\left(P^2+\varepsilon^2_f \right)^3} \frac{\Delta^2 \tilde{T}(\Delta)}{12 m^2}
    \label{F2IPMVmodel} \,.
\end{align}
Similarly, one has
\begin{align}
    \partial_P \tilde{G}_2(P,\Delta) & = - 2\pi Q_s^2  \frac{  P(P^2-\varepsilon_f^2)}{\varepsilon_f \left(P^2+\varepsilon^2_f \right)^3} \frac{\Delta^2 \tilde{T}(\Delta)}{12 m^2} 
    \label{G2IPMVmodel} \,.
\end{align}
The expressions for $\tilde{F}_0$ and $\partial_P \tilde{G}_0$, on the other hand, cannot be solved exactly due to the presence of the logarithm in the $r$ dependent part of the integrand. For example, one has
\begin{align}
     \tilde{F}_0(P,\Delta) =& \frac{\pi}{2} Q^2_s \tilde{T}(\Delta) \int r dr J_0(Pr) f_0(r) \,,
     \label{F0IPMVintermediate}
\end{align}
with
\begin{align}
    f_0(r) = r^2 \log(\frac{1}{m^2 r^2}+e) K_0(\varepsilon_f r) \,. \label{f0IPMV}
\end{align}
It would be useful to approximate Eq.\,\eqref{f0IPMV} by an expression of the form of $r^2 K_0(\varepsilon_f r)$ as it appears in the GBW model, for which we had an analytic solution. First, one should note that the convolution (Fourier transform) in Eq.\,\eqref{F0IPMVintermediate} is dominated by the maximum of $f_0(r)$. Thus, in the following we focus on reproducing the effect of the modified location and height of the maximum of Eq.\,\eqref{f0IPMV}.

Because of the logarithmic factor, Eq.\,\eqref{f0IPMV} develops a maximum at a smaller value of $r$ compared to $r^2 K_0(\varepsilon_f r)$, which depends on the ratio $\kappa \equiv \varepsilon_f/m$. We will assume that $\kappa \gg 1$. To see this more explicitly, we change to the variable $u= \varepsilon_f r$
\begin{align}
    f_0(u) = \frac{u^2}{\varepsilon^2_f} \log(\frac{\kappa^2}{u^2}+e)K_0(u) \,.
\end{align}
The maximum of this function occurs at (ignoring the factor of $e$ inside the logarithm)
\begin{align}
    u_{max} = \left[ \frac{\log(\kappa^2/u_{max}^2)-1}{\log(\kappa^2/u_{max}^2)} \right] \frac{2 K_0(u_{max})}{K_1(u_{max})} \,,
\end{align}
while in the GBW model the maximum occurs at $u_{max} =  2 K_0(u_{max})/K_1(u_{max}) \approx 1.5$.

Therefore, we see that in the IP-MV model, the location of the maximum is shifted to a smaller value of $u$ (compared to GBW):
\begin{align}
    u_{max} \approx  1.5 / \xi \,,
\end{align}
with $\xi = \left[ \frac{\log(\kappa^2)}{\log(\kappa^2)-1} \right]$. For values of $\kappa=3-10$, one has $\xi$= 1.3 - 1.8.

The corresponding maximum of $f_0$ is then
\begin{align}
   f_0(u_{max} ) \approx \frac{1.5^2 }{\xi^2 \varepsilon^2_f} \log(\kappa^2)K_0(1.5 /\xi) \,.
\end{align}
We thus approximate $f_0$ in Eq.(\ref{f0IPMV}) by
\begin{align}
    f_0(r) \approx C_1  r^2  K_0(\xi \varepsilon_f r) \,,
\end{align}
where $C_1 = \log(\kappa^2) K_0(1.5/\xi)/K_0(1.5) > 1$.

This expression reflects the shift in the location of the maximum and the increase in the height of the maximum.

Using this expression in Eq.\,\eqref{F0IPMVintermediate}, we arrive at
\begin{align}
    \tilde{F}_0(P,\Delta) & \approx  -2\pi C_1 Q_s^2 \frac{  \left(P^2 - \xi^2 \varepsilon^2_f \right)}{\left(P^2+\xi^2\varepsilon^2_f \right)^3} \tilde{T}(\Delta) \,.
    \label{F0IPMVmodel}
\end{align}
Similarly, one can approximate
\begin{align}
    \tilde{G}_0(P,\Delta)  \approx -2\pi C_2  Q_s^2  \frac{ 2  \xi P \varepsilon_f  }{\left(P^2+\xi^2\varepsilon^2_f \right)^3} \tilde{T}(\Delta)
    \label{G0IPMVmodel} \,,
\end{align}
where $C_2 = \log(\kappa^2) K_1(1.5/\xi)/K_1(1.5) > 1$.

\section{Dipole amplitude in the impact parameter dependent McLerran Venugopalan model}
\label{app:ipmv}

We briefly summarize the derivation of the dipole expressions in Eqs.\,\eqref{IPMVnoangular} and \eqref{IPMVangular}. More details on these calculations can be found in \cite{Iancu:2017fzn}.

As described in Section \ref{sec:cohdiffdipoles}, large-$x$ partons are treated as static color charges $\rho^a$ that produce color fields $A^{a,\mu}$ via Yang-Mills equations. These color fields represent the small$-x$ partons. In the Impact parameter dependent McLerran Venugopalan model, the distribution of color charges $\rho^a$ are described by local (in coordinate space and color space) Gaussian distributions
\begin{align}
\left \langle \rho^a(\bm{x}_0)\rho^b(\bm{x}_1) \right \rangle = g^2 \mu^2 \delta^{ab} \delta^{(2)}(\bm{x}_0 - \bm{x}_1) T(\bm{x}_0) \,, \label{CorrelatorSources}
\end{align}
where $T(\bm{x})$ is the transverse profile of color charges carrying the impact parameter dependence.

In the covariant gauge $(\partial_\mu A^{\mu}=0)$, the gauge fields have the form $A^{a,\mu} = \delta^{\mu +} \alpha^a$, where $\alpha^a$ satisfied the 2D Poisson equation
\begin{align} \label{Poissonalpha}
    (\bm{\nabla}^2 -m^2)\alpha^a (\bm{x})= -\rho^a(\bm{x}) \,,
\end{align} 
where the ``gluon mass'' $m$ is introduced to mimic confinement.

From Eqs.\,\eqref{CorrelatorSources} and \eqref{Poissonalpha}, one can find that the correlator of $\alpha^a$'s is given by
\begin{align}
    \left \langle \alpha^a(\bm{x}_0)\alpha^b(\bm{x}_1) \right \rangle =  \delta^{ab} \gamma(\bm{x}_0,\bm{x}_1) \,, \label{Correlatoralpha}
\end{align}
where 
\begin{align}
    \gamma({\bm{x}_0,\bm{x}_1}) = \int \frac{d^2 \bm{k}_0}{(2\pi)^2} \frac{d^2 \bm{k}_1}{(2\pi)^2} & \frac{e^{i\bm{k}_0\cdot \bm{x}_0 }}{k_0^2+m^2} \frac{e^{i\bm{k}_1\cdot \bm{x}_1}}{k_1^2+m^2} \nonumber \\ & g^2\mu^2  \tilde{T}(\bm{k}_0+\bm{k}_1) \,.
\end{align}
From the definition of the longitudinal Wilson line (Eq.\,\eqref{WilsonLine}) and the correlator above, one finds
\begin{align}
    \left \langle V^\dagger(\bm{x}_0)V(\bm{x}_1) \right \rangle =  e^{-\mathcal{N}(\bm{x}_0,\bm{x}_1)} \,, \label{CorrelatorWilson}
\end{align}
where
\begin{align}
    \mathcal{N}(\bm{x}_0,\bm{x}_1) = \frac{ g^4 \mu^2 C_F}{2}&\int  \frac{d^2 \bm{k}_0}{(2\pi)^2} \frac{d^2 \bm{k}_1}{(2\pi)^2}  \frac{(e^{i\bm{k}_0\cdot \bm{x}_0} - e^{i\bm{k}_0\cdot \bm{x}_1} )}{k_0^2+m^2} \nonumber \\ & \times \frac{(e^{i\bm{k}_1\cdot \bm{x}_0}-e^{i\bm{k}_1\cdot \bm{x}_1})}{k_1^2+m^2}  \tilde{T}(\bm{k}_0+\bm{k}_1) \,.
\end{align}
or in the convenient choice of coordinates of Eq.\,\eqref{rbcoordinates}, we have
\begin{align}
    \mathcal{N}(\bm{r},\bm{b}) = \frac{ g^4 \mu^2 C_F}{2}&\int  \frac{d^2 \bm{q}}{(2\pi)^2} \frac{d^2 \bm{k}}{(2\pi)^2}  \frac{\tilde{T}(\bm{q}) e^{i\bm{q} \cdot \bm{b}} }{(\bm{k}+\bm{q}/2)^2+m^2} \nonumber \\ & \times \frac{(e^{i \bm{q} \cdot \bm{r}/2} + e^{-i \bm{q} \cdot \bm{r}/2}-2e^{i \bm{k} \cdot \bm{r}})}{(\bm{k}-\bm{q}/2)^2+m^2}   \,.
\end{align}
This integral will be dominated by values $k \sim q \sim 1/R$. If one is interested in dipole sizes much smaller than the scale controlling the variation of the target: $r \ll R$, then one can expand the oscillating exponents in the second bracket:
\begin{align}
e^{i \bm{q} \cdot \bm{r}/2} + e^{-i \bm{q} \cdot \bm{r}/2}-2e^{i \bm{k} \cdot \bm{r}} \approx &-2i (\bm{k}\cdot \bm{r})+ (\bm{k} \cdot \bm{r})^2 \nonumber \\& - \frac{1}{4} (\bm{q}\cdot \bm{r})^2 + ...
\end{align}
Then one has
\begin{align}
    \mathcal{N}(\bm{r},\bm{b}) \approx \frac{g^4 C_F}{2} r^i r^j \int & \frac{d^2 \bm{q}}{(2\pi)^2} \frac{d^2 \bm{k}}{(2\pi)^2}  \frac{\tilde{T}(\bm{q}) e^{i\bm{q} \cdot \bm{b}} }{(\bm{k}+\bm{q}/2)^2+m^2} \nonumber \\ & \frac{(k^i k^j - q^i q^j/4)}{(\bm{k}-\bm{q}/2)^2+m^2} \,.
\end{align}
The double integral has the tensorial structure involving $\delta^{ij}$ and $2 b^ib^j/b^2 - \delta^{ij}$ (orthogonal tensors) which allows for the expansion
\begin{align}
    \mathcal{N}(\bm{r},\bm{b}) = \mathcal{N}_0(r,b) + \mathcal{N}_2(r,b) \cos 2\theta_{rb} \,,
\end{align}
where
\begin{align}
    \mathcal{N}_0(r,b) &= \frac{1}{4}Q_s^2r^2 T(b) \log(\frac{1}{r^2m^2}+e) + ...\,, \nonumber \\
    \mathcal{N}_2(r,b) &= \frac{1}{4}Q_s^2r^2\frac{1}{\pi}\int q dq \tilde{T}(q)J_2(qb) \Theta(q,m) \,, \label{DipoleEdmondRezaeian}
\end{align}
with $Q^2_s = \frac{C_F g^4 \mu^2}{4 \pi}$ and
\begin{align}
\Theta(q,m) = \int\limits_{0}^{\infty} kdk &\left[ \frac{\mathcal{I}}{(\mathcal{I}+m^2 )\sqrt{(\mathcal{I}+m^2)^2 -k^2 q^2 }} \right. \nonumber \\ &\left. +\frac{2}{q^2} - \frac{2(\mathcal{I}+m^2 )}{q^2 \sqrt{(\mathcal{I}+m^2)^2 -k^2 q^2}} \right] \,, \nonumber \\
\mathcal{I}(k,q) = k^2 + &q^2/4 \,. \label{ThetaKernel}
\end{align}
The integral in Eq.\,\eqref{ThetaKernel} results in
\begin{align}
    \Theta(q,m) = \frac{1}{2}  \left[1 -\frac{\sinh^{-1} \frac{q}{2m}}{\frac{q}{2m}\sqrt{1 + \left( \frac{q}{2m} \right)^2 } } \right] \,.
\end{align}
If one expands in powers of $q/2m$, one finds
\begin{align} \label{Thetakernelexpansion}
    \Theta(q,m) = \frac{1}{3}\left(\frac{q}{2m} \right)^2 + ...
\end{align}
Replacing this expression in Eq.\,\eqref{DipoleEdmondRezaeian} one obtains 
\begin{align} \label{N2dipoleintermediate}
    \mathcal{N}_2(r,b) &= \frac{1}{4}Q_s^2r^2 \frac{1}{6 m^2}\frac{1}{2\pi}\int q dq q^2 \tilde{T}(q)J_2(qb) \,.
\end{align}
Using the identity in Eq.\,\eqref{IdentityJ2} one obtains the dipole form in Eq.\,\eqref{IPMVangular}.
The validity of this expansion can be understood as follows: the dipole receives small momentum transfer kicks $q \sim 1/R$ with each scattering. The approximation above then is valid if $1/R < 2m$. For a large target such as a nucleus this is satisfied, while for a proton the approximation is questionable. Since the single scattering momentum transfers are restricted to $q \lesssim 2m$, we will not trust this approximation much beyond $\Delta \sim 2m$.
Even though one might be concerned about the divergence of  as $m \rightarrow 0$ in Eq.\,\eqref{IPMVangular}, one should note that in the $\lim_{m \rightarrow 0} \Theta(q,m) =  1/2$; whose effect is to replace the regulator $m$ by the finite system size, or $1/R$ in Eq.\,\eqref{IPMVangular}.

\section{Nuclear saturation scale}
\label{app:QsNuclear}

The local saturation scale $Q^2_s(b) = Q^2_{s} T(b)$ is proportional to the charge density squared of the target at point $b$. Since the nucleons are assumed to be uncorrelated, the total charge squared in the nucleus is the sum of the charges squared of all its nucleons. Thus, one has
\begin{align}
    Q^2_{sA} \int d^2 \bm{b} T_A(\bm{b}) = A Q^2_{sp} \int d^2 \bm{b} T_p(\bm{b}) \,.
\end{align}
For our choice of proton profile Eq.\,\eqref{protonprofile} and nuclear profile Eq.\,\eqref{nuclearprofile}, and using
\begin{align}
   \int d^2 \bm{b} T_p(\bm{b}) = 2 \pi R_p^2 \nonumber \\ 
   \int d^2 \bm{b} T_A(\bm{b}) \approx \frac{2\pi R_A^2}{3} \,,
\end{align}
where we assumed $R_A \gg a_A$ to approximate the $\rho_A$ in Eq.\,\eqref{nucleardensity} by a hard sphere and $N_A \approx 1/(2R_A)$. Thus we have
\begin{align}
    Q^2_{sA} = 3A \left(\frac{R_p}{R_A}\right)^2 Q^2_{sp} \approx 0.4   A^{1/3} Q^2_{sp} \,,
\end{align}
where we used $R_p =0.4 $ fm and the approximate expression $R_A = 1.1 A^{1/3}$ fm for large nuclei. A similar expression was obtained in \cite{Kowalski:2007rw}, where the authors assumed a cylindrical shape for nuclei. An expression assuming spherical nuclei and nucleons was obtained in \cite{Kovchegov:1996ty}.

If one accounts for the non-zero $a_A$, then one finds the following relation between the saturation scales: $Q^2_{sAu} = 2.17 \, Q^2_{sp}$ and $Q^2_{sCu} =  1.44 \, Q^2_{sp}$, for gold and copper respectively.

\vspace{5cm}

\bibliography{references}

\end{document}